\documentclass[aps,prd,preprint,superscriptaddress,showpacs]{revtex4}
\usepackage{graphicx}
\usepackage{float}
\usepackage{ulem}
\usepackage{color}
\definecolor{My_Red}        {cmyk}{0.00,1.00,1.00,0.20}

\newcommand{\bmat}{\left(\begin{array}}
\newcommand{\emat}{\end{array}\right)}
\newcommand{\beq}{\begin{equation}}
\newcommand{\eeq}{\end{equation}}

\def\bwt{\begin{widetext}}
\def\ewt{\end{widetext}}
\def\be{\begin{equation}}
\def\ee{\end{equation}}
\def\bea{\begin{eqnarray}}
\def\eea{\end{eqnarray}}
\def\bean{\begin{eqnarray*}}
\def\eean{\end{eqnarray*}}
\def\bary{\begin{array}}
\def\eary{\end{array}}
\def\bit{\begin{itemize}}
\def\eit{\end{itemize}}

\def\su5u1{SU(5) \times U(1)}
\def\fsu5u1{SU(5) \times U(1)'}
\def\so10{SO(10)}
\def\sq20{SO(10) \times SO(10)}

\def\bwt{\begin{widetext}}
\def\ewt{\end{widetext}}
\def\be{\begin{equation}}
\def\ee{\end{equation}}
\def\bea{\begin{eqnarray}}
\def\eea{\end{eqnarray}}
\def\bean{\begin{eqnarray*}}
\def\eean{\end{eqnarray*}}
\def\bary{\begin{array}}
\def\eary{\end{array}}
\def\bit{\begin{itemize}}
\def\eit{\end{itemize}}

\def\su5u1{SU(5) \times U(1)}
\def\fsu5u1{SU(5) \times U(1)'}
\def\so10{SO(10)}
\def\sq20{SO(10) \times SO(10)}

\usepackage[centertags]{amsmath}
\usepackage{amssymb}
\usepackage[pdftex]{hyperref}

\begin{document}

\title{Naturalness, dark matter, and the muon anomalous magnetic moment in
	supersymmetric extensions of the standard model with a pseudo-Dirac
	gluino}

\author{Chuang Li}
\affiliation{
Key Laboratory of Theoretical Physics, Institute of Theoretical Physics,
Chinese Academy of Sciences, Beijing 100190, China
}

\author{Bin Zhu}

\affiliation{Department of Physics, Yantai University, Yantai 264005, P. R. China}

\author{Tianjun Li}
\affiliation{
Key Laboratory of Theoretical Physics, Institute of Theoretical Physics,
Chinese Academy of Sciences, Beijing 100190, China
}
\affiliation{
School of Physical Sciences, University of Chinese Academy of Sciences,
No.~19A Yuquan Road, Beijing 100049, China
}

\date{\today}

\begin{abstract}

We study the naturalness, dark matter, and muon anomalous magnetic moment in the Supersymmetric Standard Models (SSMs) with a pseudo-Dirac gluino (PDGSSMs) from hybrid $F-$ and $D-$term supersymmetry (SUSY) breakings. To obtain the observed dark matter relic density and explain the muon anomalous magnetic moment, we find that the low energy fine-tuning measures are larger than about 30 due to strong constraints from the LUX and PANDAX experiments.
Thus, to study the natural PDGSSMs, we consider multi-component dark matter and then the relic density of the lighest supersymmetric particle (LSP) neutralino is smaller than the correct value. We classify our models into six kinds:
(i) Case A is a general case, which has small low energy fine-tuning measure and can explain the anomalous magnetic moment of the muon;
(ii) Case B with the LSP neutralino and light stau coannihilation;
(iii) Case C with Higgs funnel;
(iv) Case D with Higgsino LSP;
(v) Case E with light stau coannihilation and Higgsino LSP;
(vi) Case F with Higgs funnel and Higgsino LSP. We study these Cases in details, and show that our models can be natural and consistent with the LUX and PANDAX experiments, as well as explain the muon anomalous magnetic moment. In particular, all these cases except the stau coannihilation can even have low energy fine-tuning measures around 10.

\end{abstract}

\maketitle

\section{Introduction}

Supersymmetry (SUSY) provides a natural solution to the
gauge hierarchy problem in the Standard Model (SM).
In the supersymmetric SMs (SSMs)
with $R$-parity, gauge coupling unification can be achieved, the Lightest
Supersymmetric Particle (LSP) serves as a viable dark matter (DM) candidate,
and electroweak (EW) gauge symmetry can be broken radiatively because of the
large top quark Yukawa coupling, etc.
On the other hand, gauge coupling unification
strongly suggests Grand Unified Theories (GUTs), which can be
realized from superstring theory. Thus, supersymmetry is an important bridge
between the most promising new physics beyond the SM and the high-energy fundamental physics.

It is well-known that a SM-like Higgs boson $(h)$ with
mass $m_h=125.09\pm 0.24$~GeV was discovered at the LHC~\cite{ATLAS, CMS, moriond2013}.
However, to obtain
such a SM-like Higgs boson mass in the Minimal SSM (MSSM), we need either the multi-TeV top squarks with
small mixing or TeV-scale top squarks with large mixing~\cite{Carena:2011aa, Feng:2013ena}. Also,
the LHC SUSY searches give stringent constraints on the viable parameter space of
the SSMs~\cite{LHC-SUSY}.
For example, the latest SUSY search bounds show that the gluino (${\tilde g}$) mass is
heavier than about 1.6-2.0 TeV, whereas the light stop (${\tilde t}_1$) mass is heavier than
about 800-1000 GeV (Assuming these particles' masses are well separated
from the LSP mass, if not, the bound may not be so stringent).
Thus, the big challenge is how to construct the natural SSMs, which can realize the correct Higgs boson mass, solve the SUSY electroweak fine-tuning problem, and evade the LHC SUSY search constraints.
On the other hand, the dark matter direct detection
experiments such as XENON100~\cite{Aprile:2016swn}, LUX~ \cite{Akerib:2016vxi},
 and PANDAX~ \cite{Tan:2016zwf} experiments have given strong constraints
on the dark matter-nucleon spin-independent scattering cross section.
Moreover, the anomalous magnetic moment of the muon $a_{\mu}=(g_{\mu}-2)/2$
is still one of strong hints for new physics since it is deviated from the SM prediction
 more than $3\sigma$ level. The discrepancy compared to its SM theoretical value
is~\cite{Bennett:2006fi, Bennett:2008dy, Davier:2010nc}
\begin{align}
\Delta a_{\mu}=(a_{\mu})_{\text{exp}}-(a_{\mu})_{\text{SM}}
=(28.6\pm 8.0)\times 10^{-10}~.
\end{align}
In the SSMs, the light smuon, muon-sneutrino, Bino, Winos, and Higgsinos would contribute
to $\Delta a_{\mu}$~\cite{Moroi:1995yh,Martin:2001st,Byrne:2002cw,Stockinger:2006zn,Domingo:2008bb, Queiroz:2014zfa}.
Their contributions to $\Delta a_{\mu}$ from the neutralino-smuon and chargino-sneutrino loops can approximately be
expressed as
\begin{align}
 \Delta a_{\mu}^{\tilde \chi^0 \tilde \mu} &\simeq \frac{1}{192\pi^2} \frac{m_{\mu}^2}{M_{SUSY}^2} \left({\rm sgn(\mu M_1)}g_1^2-{\rm sgn(\mu M_2)}g_2^2\right)\tan \beta~, \\
 \Delta a_{\mu}^{\tilde \chi^{\pm} \tilde \nu} &\simeq {\rm sgn(\mu M_2)}\frac{1}{32\pi^2} \frac{m_{\mu}^2}{M_{SUSY}^2} g_2^2 \tan \beta~.
 \end{align}
where $M_{SUSY}$ is the typical mass scale of relevant sparticles.
 Obviously, if all the relevant sparticles have masses around the same scale, the chargino-sneutrino loop
contributions would be dominant. Thus, we have
$\Delta a_{\mu} \sim 10^{-9} \left(\frac{100~{\rm GeV}}{M_{\rm SUSY}}\right)^2 \tan \beta$ for
sgn$(\mu M_2)>0$. From Ref.~\cite{Byrne:2002cw},
we obtain that the 2$\sigma$ bound on $\Delta a_{\mu}$ can
be achieved for $\tan \beta=10$ if four relevant
sparticles are lighter than $600-700$ GeV.
While for smaller $\tan \beta$ ($\sim$3), the lighter sparticles ($\lesssim 500$ GeV) are needed.
Therefore, to explain the muon anomalous magnetic moment, we do need
the light smuon, muon-sneutrino, Bino, Winos, and Higgsinos.

Note that all the sparticles except gluino in the SSMs can be within about 1 TeV as long as
the gluino is heavier than 3 TeV, which is clearly an simple modification to
the SSMs before the LHC, we have proposed the SSMs with a pseudo-Dirac
gluino (PDGSSMs) from hybrid $F-$ and $D-$term SUSY breakings, which
can explain the dark matter and muon anomalous magnetic moment simultaneously~\cite{Ding:2015wma}.
Such kind of models solves the following problems in the SSMs
with Dirac gauginos or say supersoft SUSY~\cite{Fox:2002bu,Benakli:2008pg,Benakli:2010gi,Kribs:2012gx,Benakli:2012cy,Kribs:2013oda,Bertuzzo:2014bwa,Benakli:2014cia,Diessner:2014ksa, Nelson:2015cea,Kribs:2013eua,diCortona:2016fsn} due to the $F-$term gravity mediation:
$\mu$ problem cannot be solved via the Giudice-Masiero (GM)
mechanism~\cite{Giudice:1988yz}, the D-term contribution
to the Higgs quartic coupling vanishes, and the right-handed slepton
may be the LSP, etc~\cite{Fox:2002bu}. There is another problem
in supersoft SUSY: the scalar components of the adjoint chiral superfields may be tachyonic and then break the SM gauge symmetry,
which was  solved in Ref.~\cite{Nelson:2015cea}.
In the PDGSSMs, all the sparticles in the MSSM obtain the SUSY breaking soft terms from the traditional gravity mediation,
while only gluino receives extra Dirac mass from the $D-$term SUSY breaking.
In particular, all the MSSM sparticles except gluino can be within about 1 TeV
as the pre-LHC SSMs. In short, we can keep the
merits of pre-LHC SSMs (naturalness, and explanations for
the dark matter and muon anomalous magnetic moment, etc),
evade the LHC SUSY search constraints, and solve
the problems in supersoft SUSY via the $F$-term gravity mediation.
We also showed that such SUSY breakings can be realized
by an anomalous $U(1)_X$ gauge symmetry inspired from string models. Moreover,
in order to obtain the gauge coupling unification and lift the Higgs boson mass,
we will introduce vector-like particles.
As a side comment, the PDGSSMs are different
from the SSMs with EW SUSY (EWSUSY)~\cite{Cheng:2012np, Cheng:2013hna, Li:2014dna, Zhu:2016ncq}
Besides the psedo-Dirac and Majorana gluinos,
the main difference is that the squarks are light in the PDGSSMs while
heavy in the SSMs with EWSUSY. 
For an overview about naturalness or fine-tuning in the SSMs, please see ~\cite{Arvanitaki:2013yja}. 
We have solved the relevant problems mentioned in ~\cite{Arvanitaki:2013yja}, and concentrate on low energy phenomenology(especially dark matter) here. 

In this paper, we will study the naturalness, dark matter, and muon anomalous magnetic moment
in the PDGSSMs. To obtain the observed dark matter density and explain the
muon anomalous magnetic moment, we show that the low energy fine-tuning measures are
larger than about 30 due to strong constraints from the LUX and PANDAX experiments.
Thus, to realize the natural PDGSSMs, we consider multi-component dark matter
and then the relic density of the LSP neutralino is
smaller than the correct value. We classify the dark matter models into six kinds:
(i) Case A is a general case, which has small low energy fine-tuning measure and
can explain the anomalous magnetic moment of the muon; (ii) Case B with the LSP neutralino
and light stau coannihilation; (iii) Case C with Higgs funnel; (iv) Case D with Higgsino LSP;
(v) Case E with light stau coannihilation and Higgsino LSP;
(vi) Case F with Higgs funnel and Higgsino LSP. We discuss these Cases in details,
and find that our models can be natural and consistent with the LUX and PANDAX experiments,
and explain the muon anomalous magnetic moment as well. In particular,
all these cases except the stau coannihilation can even have low energy fine-tuning
measures around 10.

\section{Brief Review of the PDGSSMs}

To obtain the Dirac gluino mass, we introduce a chiral superfield $\Phi$ in the adjoint
representation of $SU(3)_C$. To achieve the gauge coupling unification and
increase the Higgs boson mass, we introduce some extra vector-like particles.
In order to solve the Landau pole problem for the SM gauge couplings
below the GUT scale, we only have two kinds of models:
$\Delta b=3$ and $\Delta b=4$ where $\Delta b$ is the universal contribution to
the one-loop beta functions of the SM gauge couplings from all the new particles(i.e. $\beta_g = (1/16 \pi^2)(b+\Delta b) g^3$).
The additional vector-like particles and their quantum numbers in the SSMs with $\Delta b =3$
and $\Delta b =4$ are given in Tables \ref{tab:Db=3} and \ref{tab:Db=4}, respectively.
Similar to our previous paper, we still consider the model with $\Delta b=4$, while
the model with $\Delta b=3$ will be studied elsewhere (For Dirac gaugino case, see  Ref.~\cite{Benakli:2014cia}.).
In the model with $\Delta b=4$, the $SU(2)_L\times U(1)_Y$ Dirac gaugino masses are forbidden, and
the neutrino masses and mixings can be generated via Type II seesaw mechanism~\cite{Konetschny:1977bn}.

\begin{table}[h]
\begin{tabular}{|c|c|c|c|}
\hline
Particles & Quantum Numbers  & Particles  & Quantum Numbers \\
\hline
$\Phi$ & $(\mathbf{8}, \mathbf{1}, \mathbf{0})$ & $T$ & $(\mathbf{1}, \mathbf{3}, \mathbf{0})$ \\
\hline
$XL$ & $(\mathbf{1}, \mathbf{2}, \mathbf{-1/2})$ & $XL^c$ & $(\mathbf{1}, \mathbf{2}, \mathbf{1/2})$ \\
\hline
$XE_i$ & $(\mathbf{1}, \mathbf{1}, \mathbf{-1})$ & $XE^c_i$ & $(\mathbf{1}, \mathbf{1}, \mathbf{1})$ \\
\hline
$S$ & $(\mathbf{1}, \mathbf{1}, \mathbf{0})$ & & \\
\hline
\end{tabular}
\caption{The extra vector-like particles and their quantum numbers in the supersymmetric SM with $\Delta b =3$.
Here, $i=1,~2$, and we do not have to introduce $S$ except for Dirac gaugino case
since it is an SM singlet.}
\label{tab:Db=3}
\end{table}

\begin{table}[h]
\begin{tabular}{|c|c|c|c|}
\hline
Particles & Quantum Numbers  & Particles  & Quantum Numbers \\
\hline
$\Phi$ & $(\mathbf{8}, \mathbf{1}, \mathbf{0})$ &  & \\
\hline
$XD$ & $(\mathbf{3}, \mathbf{1}, \mathbf{-1/3})$ & $XD^c$ & $(\mathbf{\bar 3}, \mathbf{1}, \mathbf{1/3})$ \\
\hline
$T_+$ & $(\mathbf{1}, \mathbf{3}, \mathbf{1})$ & $T_-$ & $(\mathbf{1}, \mathbf{3}, \mathbf{-1})$ \\
\hline
\end{tabular}
\caption{The extra vector-like particles and their quantum numbers in the supersymmetric SM with $\Delta b =4$. }
\label{tab:Db=4}
\end{table}

Besides the MSSM superpotential, the new superpotential terms with universal vector-like particle mass are
\begin{align}
W&=  M_V (T_+ T_- + XD^c XD) +\lambda H_u T_- H_u + \lambda' H_d T_+ H_d \, ,\nonumber
\label{eq:DiracTMSSM}
\end{align}
where $\lambda$ and $\lambda'$ are Yukawa couplings,
and $H_u$ and $H_d$ are the Higgs doublet fields which give masses to the up-type quarks and down-type quarks/charged leptons, respectively. The $\lambda$ and $\lambda'$ terms
give the positive and negative contributions to Higgs boson mass via the non-decoupling
effects, respectively. The Higgs mass is lifted through the non-decoupling effect which
is mediated by the large mass splitting between $m_{T_+}$and $M_V$. The details can be found in Appendix. 
$m_{T_+}$ will contribute to the one-loop beta function of $m_{H_u}^2$ due to the $\lambda^{\prime}$ term,
and thus the fine-tuning becomes very large. In order to avoid such a dangerous scenario, 
$\lambda^{\prime}$ must be forbidden by some global symmetry consideration. We are left with $\lambda$ and $m_{T_+}$, 
which realize the DiracTMSSM model. The corresponding Landau pole problem for $\lambda$
strongly depends on its value at SUSY scale. The advantage of DiracTMSSM demonstrates 
that it only needs mild value of $\lambda$ to achieve the observed Higgs mass. 
Therefore, the Landau pole problem can be removed easily.
Then the corresponding SUSY breaking soft terms are
\begin{align}
-\mathcal{L}_{soft}&= B_T T_- T_+ +B_D XD^c XD +T_{\lambda} H_u T_- H_u\nonumber\\
                   & +M_{D} G \Phi + {\rm h.c.} +  \tilde \phi^\dagger m_{\tilde\phi}^2 \tilde\phi,
\end{align}
where $B_{T,D}$ are bilinear soft terms, $T_{\lambda}$ is a trilinear soft term, $m_{\tilde\phi}^2$ are soft scalar masses,
and  $M_{D}$ is the Dirac gluino mass.

To realize the hybrid $F-$ and $D-$term SUSY breakings,
we consider the anomalous $U(1)_X$ gauge symmetry
inspired from string models~\cite{Ding:2015wma},
and assume that all the SM particles and vector-like particles
are neutral under $U(1)_X$. Unlike Ref.~\cite{Dvali:1996rj},
we introduce two SM singlet fields $S$ and $S'$ with $U(1)_X$ charges $0$ and $-1$~\cite{Ding:2015wma}.
Generically, there might exist the other exotic particles $Q^X_i$ with
$U(1)_X$ charges $q^X_i$ from any real string compactification. Therefore,
the $U(1)_X$  $D$-term potential is
\begin{eqnarray}
V_D= {g_X^2 \over 2}D^2 = {g_X^2 \over 2}\left(\sum_i q^X_i|Q^X_i|^2 - |S'|^2+\xi\right)^2\, ,
\label{D-Potential}
\end{eqnarray}
where for example in the heterotic string compactification~\cite{Cvetic:1998gv},
the Fayet-Iliopoulos term is
\begin{eqnarray}
\xi = {g_X^2{\rm Tr}{ q^X} \over 384\pi^2}M_{\rm Pl}^2\, ,
\end{eqnarray}
with $M_{\rm Pl}$ the reduced Planck scale.

In order to have gravity mediation, we consider the following superpotential due to the instanton effect
which breaks $U(1)_X$ gauge symmetry~\cite{Ding:2015wma}
\begin{eqnarray}
W_{\rm Instanton} &=& M_I S S'~.~
\end{eqnarray}
This is the key difference between our model~\cite{Ding:2015wma}
and that in Ref.~\cite{Dvali:1996rj} where the superpotential
is $U(1)_X$ gauge invariant and thus one cannot obtain
the traditional gravity mediation. Minimizing the potential, we get
\begin{eqnarray}
&& \langle S \rangle = 0~,~~\langle S' \rangle^2
= \xi- {M_I^2 /g_X^2}~,~~ \langle F_{S'}\rangle = 0~,~\\
&& \langle F_{S}\rangle = M_I\sqrt{\xi- {M_I^2 /g_X^2}}~,~~
 \langle D\rangle ={M_I^2}/{g_X^2}\, .
\end{eqnarray}
where $F_{S}$ and $F_{S'}$ are the corresponding F-terms, D is the corresponding D-term. Note that $S$ is neutral under $U(1)_X$, the traditional gravity mediation can be realized
via the non-zero $F_S$~\cite{Ding:2015wma}. The Dirac mass for gluino/$\Phi$ and soft scalar
masses for $\Phi$ and $T_{+/-}$
can be generated respectively via the following operators~\cite{Nelson:2015cea}
\begin{eqnarray}
\int d^2\theta\left( \frac{{\overline D}^2 D^\alpha V'}{M_*} W_{3, \alpha} \Phi+
\frac{{\overline D}^2(D^{\alpha} V' D_{\alpha} X' )}{M_*} X''\right) ~,~\,
\end{eqnarray}
where for simplicity the coefficients are neglected, $W_{3, \alpha}$ is the field-strength superfield of SM gauge group SU(3), with $\alpha$ being the spinor index. $X'$ and $X''$ can
both be $\Phi$ as well as respectively be $T_{+/-}$ and $T_{-/+}$,
and $M_*$ can be either the reduced
Planck scale for gravity mediation or the effective messenger scale. Thus,
the Dirac mass for gluino/$\Phi$ and soft scalar
masses for $\Phi$ and $T_{+/-}$ can be about 3-5 TeV from D-term contributions~\cite{Ding:2015wma}.
By the way, the above operators ameliorate several drawbacks of the purely supersoft supersymmetry~\cite{Nelson:2015cea},
for example, the $\mu$ problem, and the vanishing Higgs quartic term problem, etc. Also,
the Majorana masses of the adjoint fermions might result in a lighter Bino from seesaw mechanism.
In particular, the new $\mu$-term can give unequal masses to the up and down type Higgs fields,
and the Higgsinos can be much heavier than the Higgs boson without fine-tuning. However, the unequal
Higgs and Higgsino masses remove some attractive features of supersoft supersymmetry as well.

To solve the Landau pole problem for gauge couplings below the GUT scale,
we require $M_V \ge 3$~TeV and $M_D \ge 3$~TeV. As we know,
there is no fine-tuning problem for  $M_D$ as large as 5 TeV due to supersoft supersymmetry~\cite{Kribs:2012gx}. 
However, large $M_D$ will lead to instability in numerical codes such as SPheno. Therefore, 
we choose $M_D=3$ TeV which can not only escape the LHC supersymmetry search constraints but also 
not introduce any fine-tuning issue and instability. Because the vector-like particles are introduced
to retain gauge coupling unification,  $M_V$ must be around $M_D$ as well.
The Higgs boson mass is increased via a non-decoupling effect~\cite{Ding:2015wma}
as in the Dirac NMSSM~\cite{Lu:2013cta,Kaminska:2014wia}
\begin{eqnarray}
\Delta m^2_h = 2 v^2 \lambda_{\text{eff}}^2 \sin^4\beta \,\,,
\end{eqnarray}
where $\tan\beta \equiv \langle H_u \rangle/\langle H_d \rangle$, and
\begin{equation}
\lambda_{\text{eff}}^2 \equiv \lambda^2(m_{T_+}^2/(M_{V}^2+m_{T_+}^2)) \, .
\end{equation}
Unlike the Dirac NMSSM, such contribution does not vanish at large $\tan\beta$ limit,
which is very important to explain the muon anomalous magentic moment~\cite{Ding:2015wma}.

In this paper, we only study the simple low energy phenomenology. Thus,
we consider the low energy fine-tuning measure which is defined as follows~\cite{Baer:2012mv}
\begin{equation}
\Delta_{\text{EW}}=\frac{2}{M_Z^2}\text{max}(C_{H_d},C_{H_u},C_{\mu},C_{B_{\mu}},C_{\delta m_{H_u}^2})~,~
\label{eq:FT}
\end{equation}
where
\begin{align}
& C_{H_d}=\left|\frac{m^2_{H_d}}{\tan^2\beta-1}\right|,\, C_{H_u}=\left|\frac{m^2_{H_u}\tan^2\beta}{\tan^2\beta-1}\right|,  \\
& C_{\mu}=\left|\mu^2\right|,\,C_{B_{\mu}}=\left|B_{\mu}\right|,\\
& C_{\delta m_{H_u}^2}=\frac{(\lambda M_V)^2}{16\pi^2}\log\left(\frac{M_V^2+m^2_{T_+}}{M_V^2}\right)~.~
\end{align}
Compared to Ref.~\cite{Baer:2012mv} we have extra $C_{\delta m_{H_u}^2}$
from the triplet threshold corrections to $m_{H_u^2}$.

\section{Phenomenology Study}

In this section, we will study the naturalness, dark matter, and muon anomalous magnetic moment in the PDGSSMs numerically.
For this purpose, we have implemented this model in the Mathematica package {\tt SARAH v4.8.0}~\cite{Staub:2008uz,Staub:2009bi,Staub:2010jh,Staub:2012pb,Staub:2013tta}. {\tt SARAH v4.8.0} has been used to create a {\tt SPheno v3.3.8}~\cite{Porod:2003um,Porod:2011nf} module for the PDGSSMs to calculate the mass spectrum with good precision. The generated spectrum is transfered to {\tt micrOMEGAs v4.1}~\cite{Belanger:2014vza} to calculate dark matter relic density and direct detection cross-sections with the help of {\tt CalcHEP 3.6.25}~\cite{Belyaev:2012qa}. As a whole, we use {\tt SSP v1.2.3}~\cite{Staub:2015kfa} to do the parameter space scaning.

The null results from the SUSY searches at the LHC put severe limits on the masses of gluino and squarks~\cite{LHC-SUSY}. We consider the following low bounds on sparticle masses
\begin{enumerate}
\item $m_{\tilde q1,2}\gtrsim 1.4~{\rm TeV}$.
\item $m_{\tilde q3}\gtrsim 1.0~{\rm TeV}$.
\item $m_{\tilde g}\gtrsim 2.0~{\rm TeV}$.
\item $m_{\tilde l1,2,3,4,5,6}\gtrsim 360~{\rm GeV}$ (When the light slepton coannihilates with the LSP, it may not have such stringent constraint.).
\end{enumerate}

Before we perform the numerical analysis, let us explain our convention. We define the dimensionless parameter $l_{up}\equiv -{\lambda_{eff}}^2$. For all the input mass parameters such as $\mu,~ M_1, ~M_2, ~M_D$, $m_{\Phi}^2$, $m_{\tilde{Q},\tilde{U},\tilde{D},1\&2}$, $m_{\tilde{Q},3}$, $m_{\tilde{U},3}$, $m_{\tilde{D},3}$, $m_{\tilde{L},\tilde{E},1\&2}$, $m_{\tilde{L},3}$, $m_{\tilde{E},3}$, we choose GeV unit. While for $B_{\mu}$, its unit is ${\rm GeV}^2$. In addition, the particle masses for all the benchmark points in the following tables are in GeV unit as well.

As the preferred range for the LSP neutralino relic density, we consider the 2$\sigma$ interval combined range from Planck+WP+highL+BAO~\cite{Ade:2013zuv}
\begin{equation}
0.1153<\Omega_{CDM}h^2<0.1221~.~
\label{eq:DMConstraints}
\end{equation}
However, we find that in this case, the low energy fine-tuning measures are generically larger than about 30. The reason is that the parameter spaces, which have the correct dark matter relic density and smaller fine-tuning measures, are excluded by the LUX and PANDAX experiments~\cite{Tan:2016zwf, Bennett:2006fi}. In Tables~\ref{table:BMP_StauCoan_SingleDM} and \ref{table:BMP_AReson_SingleDM}, we present two benchmark points: one for the LSP neutralino and light stau coannihilation and the other for Higgs funnel, respectively. These benchmark points have the observed dark matter relic density and $\Delta a_{\mu}$ within $1\sigma$ range.
The fine-tuning measure for the light stau coannihilation benchmark point is about 38.5, which is still acceptable.
However, the fine-tuning measure for Higgs funnel benchmark point is about 158.2, which is a little bit too large.


\begin{table}[H]
	\small
	\centering
	\caption{The particle spectrum (in GeV) for the benchmark point of stau coannihilation.
		Here, $tan\beta = 37.8478$, $l_{up} = -0.0363711$, $\mu=400$, $B_{\mu}=2\times 10^4$, $M_1 = 147.266$, $M_2 = 400$, $M_3 = 600$, $M_D = 3,000$, $m_{\Phi}=2\times10^3$, $m_{\tilde{Q},\tilde{U},\tilde{D},1\&2}=\sqrt{2}\times10^3$, $m_{\tilde{Q},3}=m_{\tilde{U},3}=m_{\tilde{D},3}=\sqrt{1.2}\times10^3$, $m_{\tilde{L},\tilde{E},1\&2}=0.5\times 10^3$, $m_{\tilde{L},3}=0.5\times 10^3$, $m_{\tilde{E},3} = 156.701$, $\Omega_{\chi}h^2 = 0.118993$, $\sigma_{SI} = 2.92879\times10^{-46}~{\rm cm}^2$, $\Delta_{EW} = 38.4838$, $\Delta a_\mu = 2.61375\times10^{-9}$. Moreover, $O_\phi$ and $O_\sigma$ are real and imaginary components of $\Phi$. We have considered right handed sleptons for coannihilation, it’s also possible for left handed sleptons, we just give a example point here}
	\begin{tabular}{|c|c||c|c||c|c||c|c||c|c||c|c||} \hline		
		 $h$&$125.889$&$\widetilde{\chi}_{1}^{0}$&$143.538$&$\widetilde{\nu}_{\tau}$&$497.714$&$\widetilde{\tau}_{1}$&$150.723$&$\widetilde{t}_{1}$&$1354.76$&$\widetilde{b}_{1}$&$1341.03$\\\hline
		 $H$&$787.83$&$\widetilde{\chi}_{2}^{0}$&$355.473$&$\widetilde{\nu}_{\mu}$&$499.263$&$\widetilde{\mu}_{1}$&$502.686$&$\widetilde{t}_{2}$&$1362.57$&$\widetilde{b}_{2}$&$1368.96$\\\hline
		 $A$&$798.514$&$\widetilde{\chi}_{3}^{0}$&$409.601$&$\widetilde{\nu}_{e}$&$499.27$&$\widetilde{e}_{1}$&$503.43$&$\widetilde{c}_{1}$&$1648.91$&$\widetilde{s}_{1}$&$1649.14$\\\hline
		 $H^{\pm}$&$813.684$&$\widetilde{\chi}_{4}^{0}$&$470.665$&$O_\phi$&$6082.4$&$\widetilde{e}_{2}$&$506.082$&$\widetilde{u}_{1}$&$1648.91$&$\widetilde{d}_{1}$&$1649.15$\\\hline
		 $\widetilde{g}_1$&$2911.4$&$\widetilde{\chi}_{1}^{\pm}$&$355.446$&$O_\sigma$&$1445.72$&$\widetilde{\mu}_{2}$&$506.801$&$\widetilde{u}_{2}$&$1653.5$&$\widetilde{d}_{2}$&$1655.56$\\\hline
		 $\widetilde{g}_2$&$3443.67$&$\widetilde{\chi}_{2}^{\pm}$&$471.278$&$-$&$-$&$\widetilde{\tau}_{2}$&$507.578$&$\widetilde{c}_{2}$&$1653.5$&$\widetilde{s}_{2}$&$1655.56$\\\hline				 
	\end{tabular}
	\label{table:BMP_StauCoan_SingleDM}
\end{table}


\begin{table}[H]
	\small
	\centering
	\caption{The particle spectrum (in GeV) for the benchmark point of Higgs funnel.
		Here, $tan\beta = 29.$, $l_{up} = -0.0242932$, $\mu=811.11$,  $B_\mu = 12807.4$, $M_1 =266.163$, $M_2 = 400$, $M_3 = 600$, $M_D = 3,000$, $m_{\Phi}=2\times10^3$, $m_{\tilde{Q},\tilde{U},\tilde{D},1\&2}=\sqrt{2}\times10^3$, $m_{\tilde{Q},3}=m_{\tilde{U},3}=m_{\tilde{D},3}=\sqrt{1.2}\times10^3$, $m_{\tilde{L},\tilde{E},1\&2}=0.5\times 10^3$, $m_{\tilde{L},3}=0.5\times 10^3$,$m_{\tilde{E},3}=0.4\times 10^3$, $\Omega_{\chi}h^2 = 0.121222$, $\sigma_{SI} = 1.42513\times10^{-46}~{\rm cm}^2$ , $\Delta_{EW} = 158.241$, $\Delta a_\mu = 1.41094\times10^{-9}$.}
	\begin{tabular}{|c|c||c|c||c|c||c|c||c|c||c|c||} \hline		
		 $h$&$126.267$&$\widetilde{\chi}_{1}^{0}$&$262.831$&$\widetilde{\nu}_{\tau}$&$498.324$&$\widetilde{\tau}_{1}$&$382.616$&$\widetilde{t}_{1}$&$1356.82$&$\widetilde{b}_{1}$&$1335.71$\\\hline
		 $H$&$495.207$&$\widetilde{\chi}_{2}^{0}$&$409.892$&$\widetilde{\nu}_{\mu}$&$499.515$&$\widetilde{\mu}_{1}$&$502.037$&$\widetilde{t}_{2}$&$1364.22$&$\widetilde{b}_{2}$&$1378.94$\\\hline
		 $A$&$501.978$&$\widetilde{\chi}_{3}^{0}$&$815.913$&$\widetilde{\nu}_{e}$&$499.518$&$\widetilde{e}_{1}$&$503.421$&$\widetilde{c}_{1}$&$1649.01$&$\widetilde{s}_{1}$&$1649.09$\\\hline
		 $H^{\pm}$&$548.246$&$\widetilde{\chi}_{4}^{0}$&$822.8$&$O_\phi$&$6082.4$&$\widetilde{e}_{2}$&$506.277$&$\widetilde{u}_{1}$&$1649.01$&$\widetilde{d}_{1}$&$1649.11$\\\hline
		 $\widetilde{g}_1$&$2911.4$&$\widetilde{\chi}_{1}^{\pm}$&$410.07$&$O_\sigma$&$1445.72$&$\widetilde{\mu}_{2}$&$507.641$&$\widetilde{u}_{2}$&$1653.47$&$\widetilde{d}_{2}$&$1655.54$\\\hline
		 $\widetilde{g}_2$&$3443.67$&$\widetilde{\chi}_{2}^{\pm}$&$823.693$&$-$&$-$&$\widetilde{\tau}_{2}$&$520.301$&$\widetilde{c}_{2}$&$1653.47$&$\widetilde{s}_{2}$&$1655.55$\\\hline				 
	\end{tabular}
	\label{table:BMP_AReson_SingleDM}
\end{table}

Therefore, we consider the multi-component dark matter in the following, and only require
\begin{equation}
\Omega_{CDM}h^2<0.1221~.~
\label{eq:DMConstraintsMax}
\end{equation}
Other dark matter components(or candidates) can be sterile neutrino~\cite{Dodelson:1993je, Adhikari:2016bei, Abazajian:2017tcc}, axion~\cite{Holman:1982tb, Sikivie:2009fv}, etc.

In the following paper, when we mention dark matter-nucleon scattering cross section of our candidate, we mean relative scattering cross section:
\begin{equation}
\sigma_{relative}=\sigma_{SI} \times (\Omega_{\chi}h^2/\Omega_{CDM}h^2).
\label{eq:Relative_Cross_Section}
\end{equation}
Then we use relative scattering cross section to compare with PandaX/LUX data.

\begin{figure} [H]
	\begin{center}
		\includegraphics[width=0.3\textwidth]{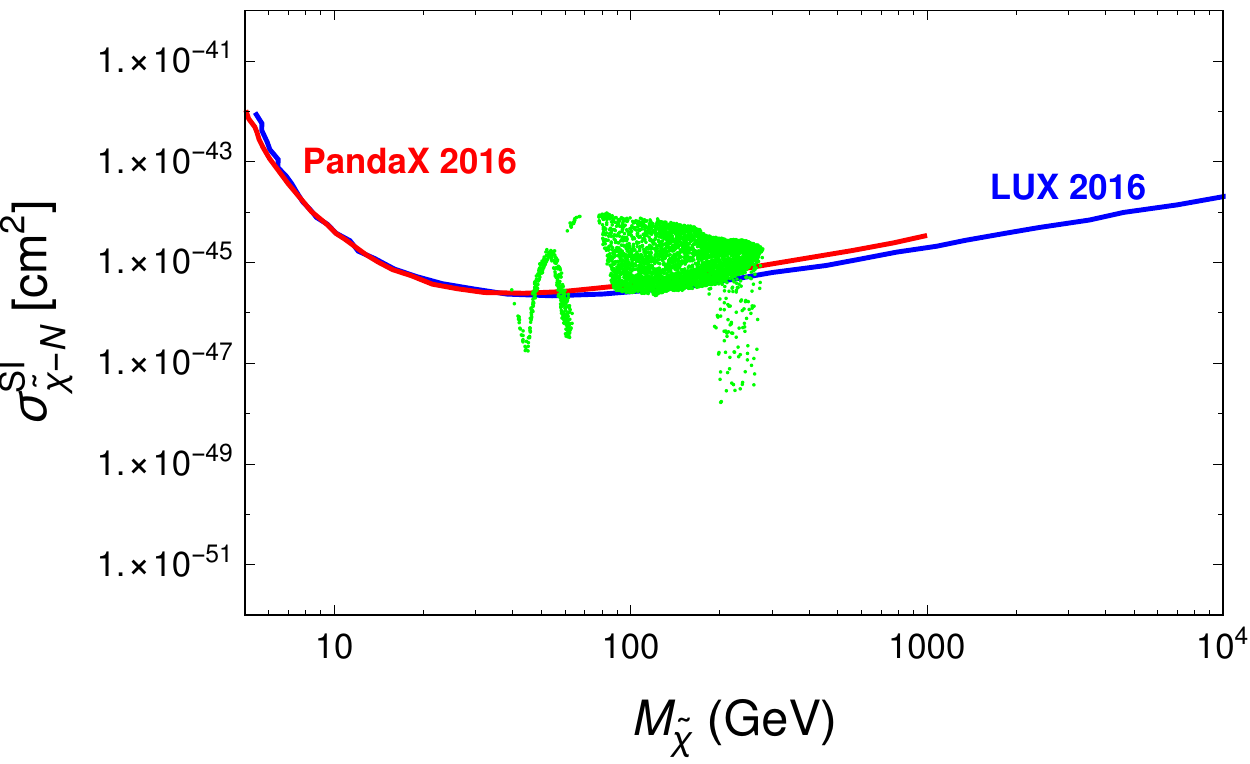}
		\includegraphics[scale=0.34]{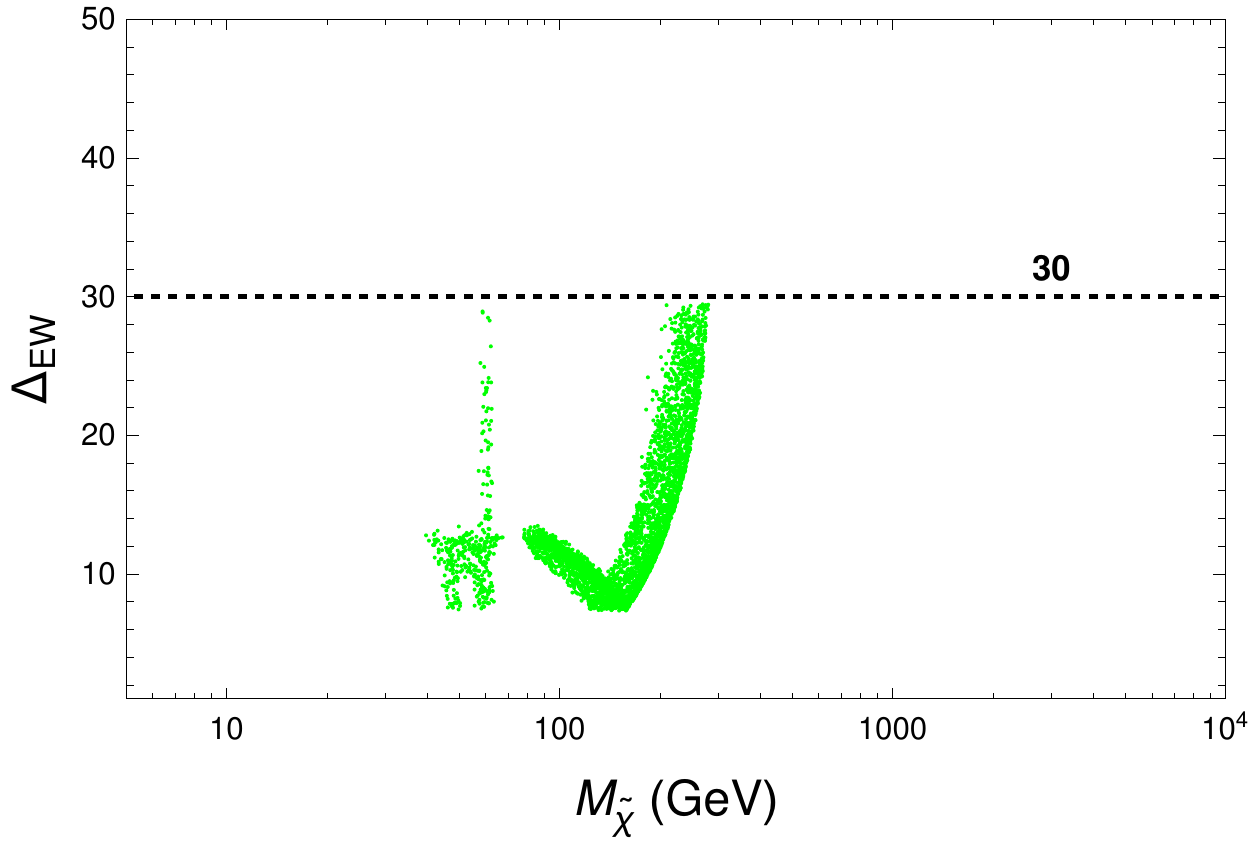}
		\includegraphics[scale=0.36]{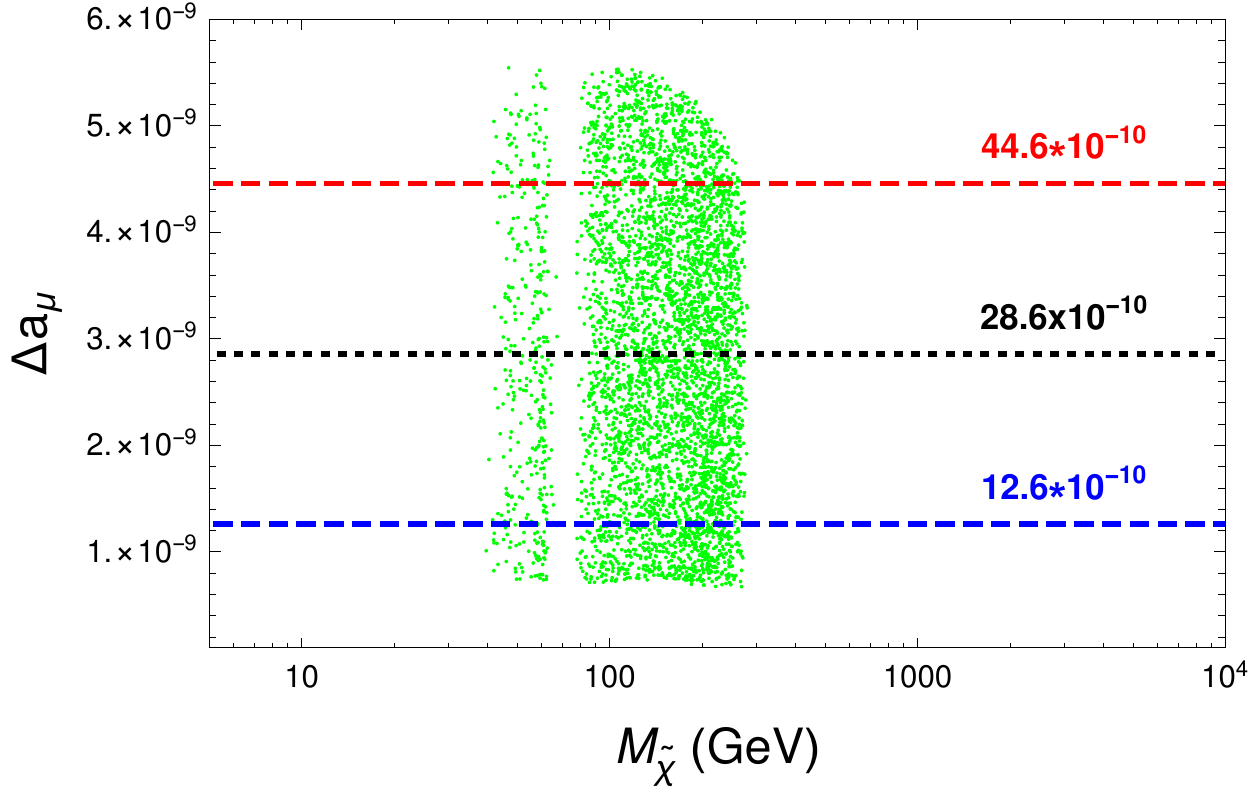}
	\end{center}
	\caption{The spin-independent elastic dark matter-nucleon scattering cross section, fine-tuning measure,
and muon anomalous magnetic moment versus the LSP neutralino mass for Case A. }
	\label{fig:GeneralScan}
\end{figure}

\begin{figure} [H]
\begin{center}
\includegraphics[width=0.3\textwidth]{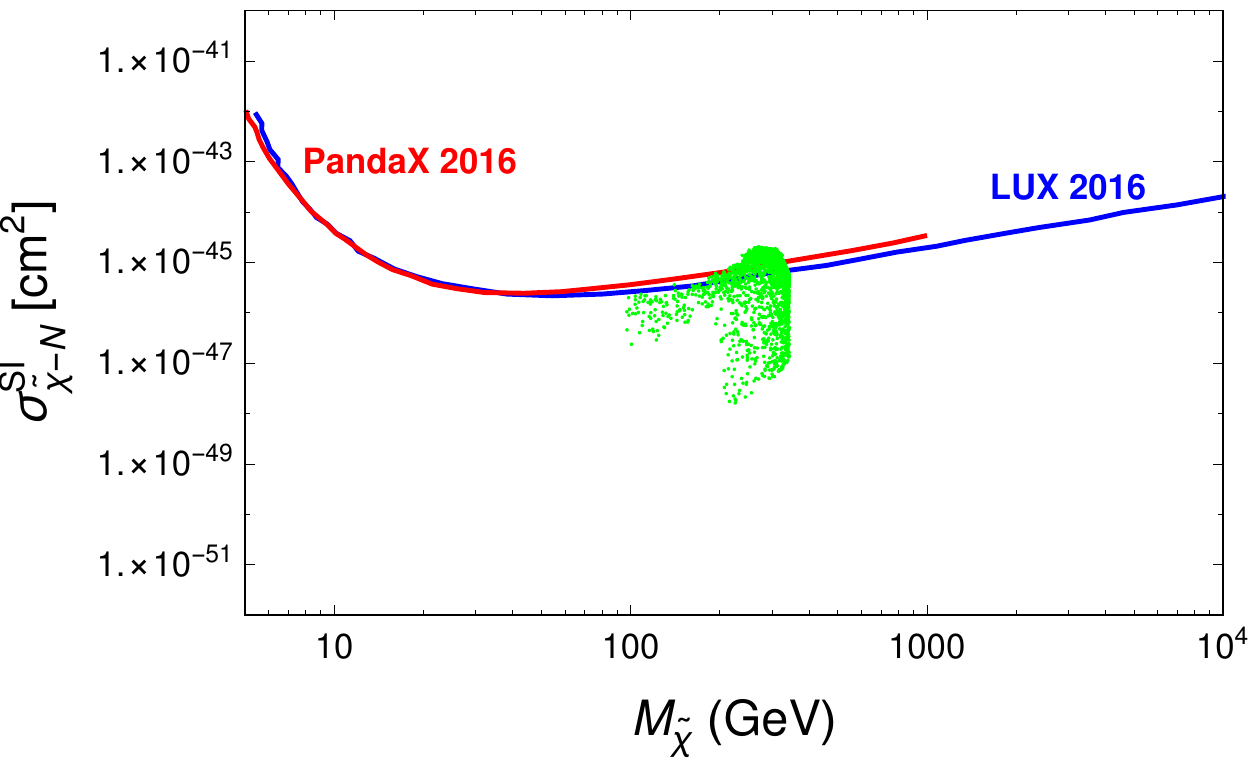}
\includegraphics[scale=0.34]{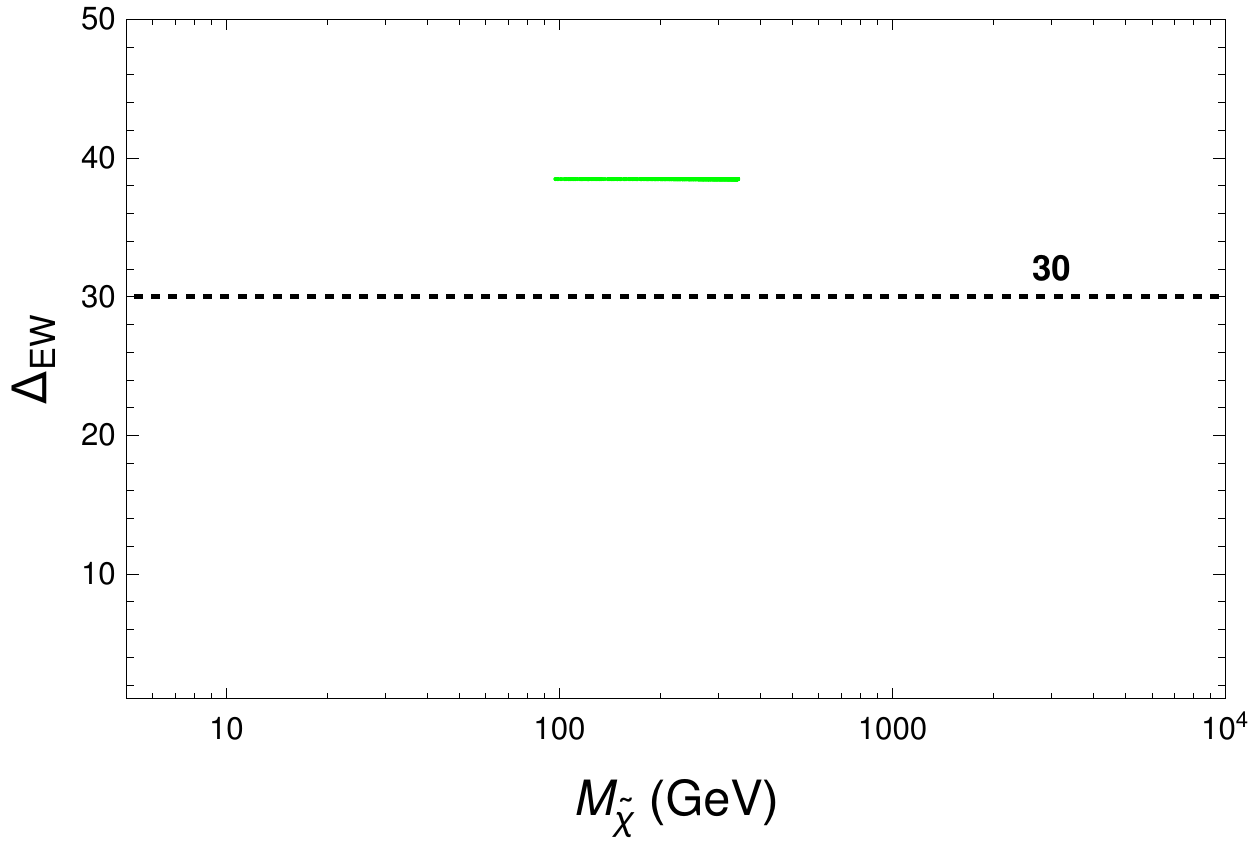}
\includegraphics[scale=0.36]{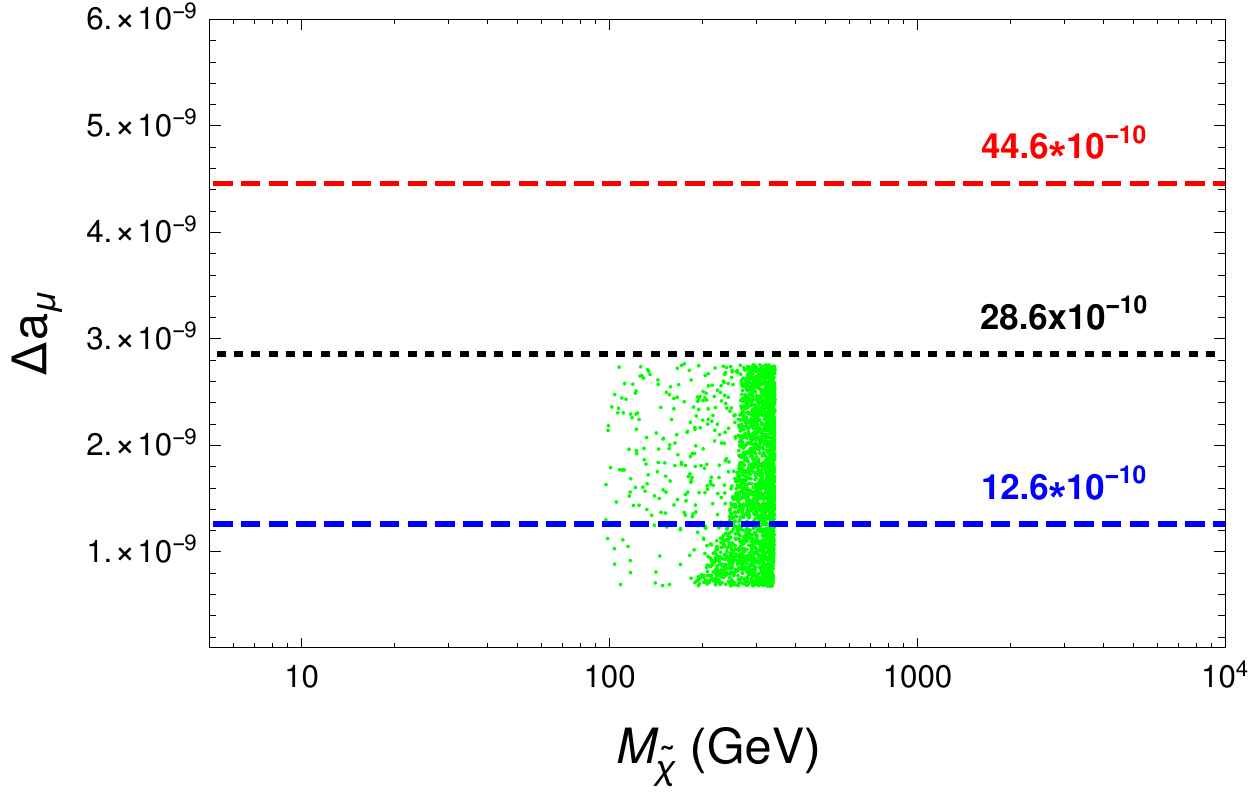}
\end{center}
\caption{The spin-independent elastic dark matter-nucleon scattering cross section, fine-tuning measure,
and muon anomalous magnetic moment versus the LSP neutralino mass for Case B.}
\label{fig:StauCoan}
\end{figure}

\begin{figure} [H]
\begin{center}
\includegraphics[width=0.3\textwidth]{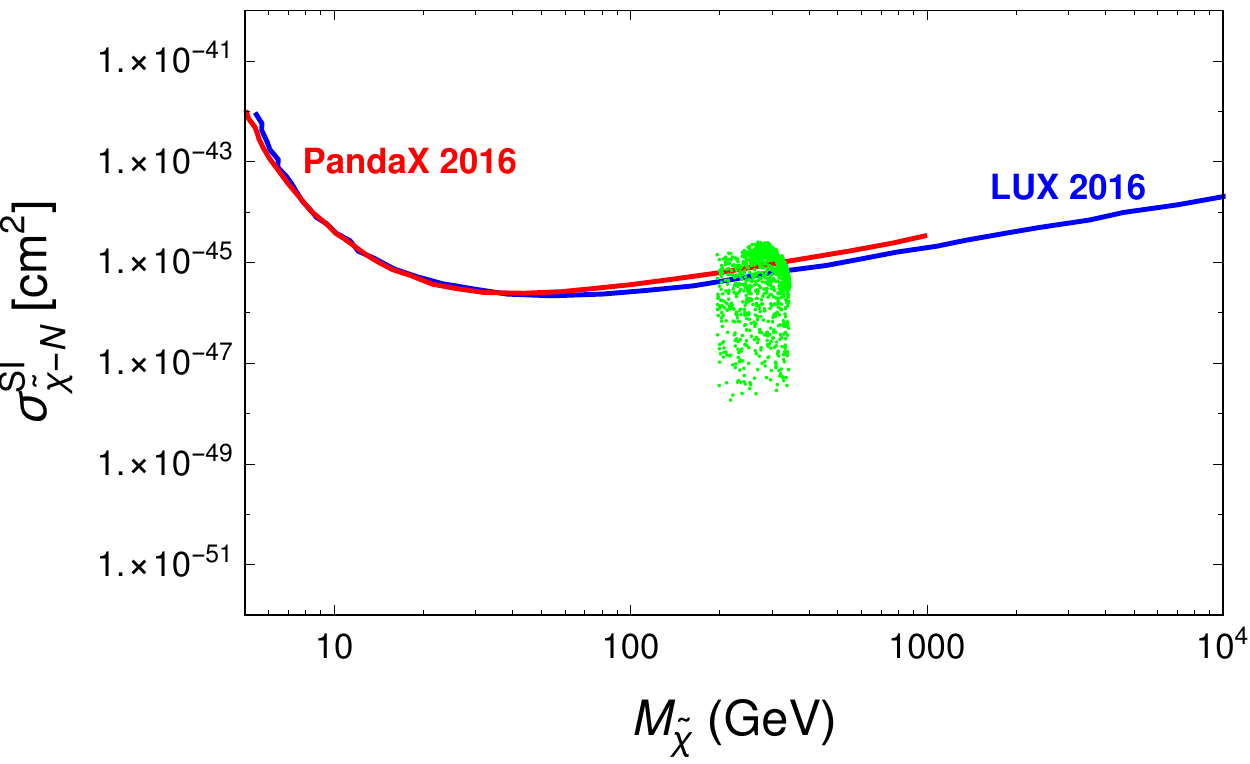}
\includegraphics[scale=0.34]{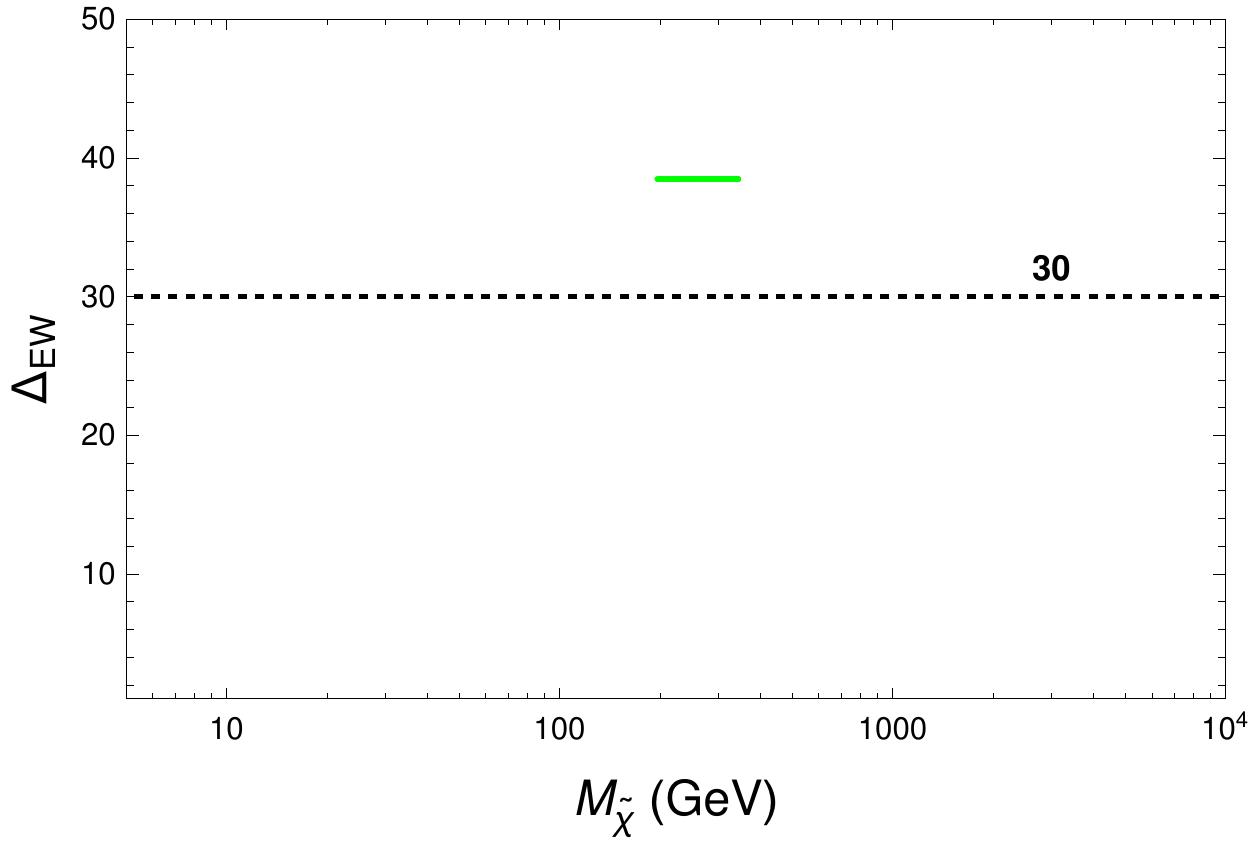}
\includegraphics[scale=0.36]{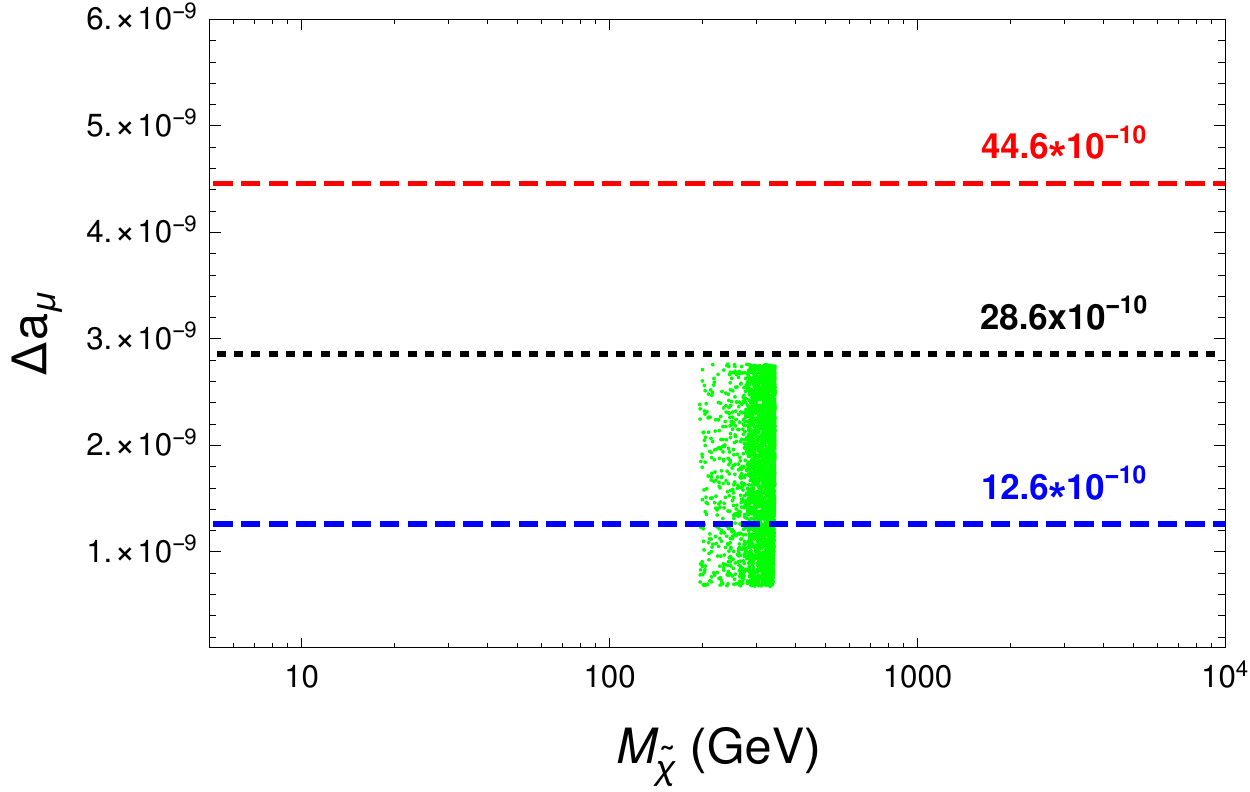}
\end{center}
\caption{The spin-independent elastic dark matter-nucleon scattering cross section, fine-tuning measure,
and muon anomalous magnetic moment versus the LSP neutralino mass for Case C.}
\label{fig:AReson}
\end{figure}

\begin{figure} [H]
\begin{center}
\includegraphics[width=0.3\textwidth]{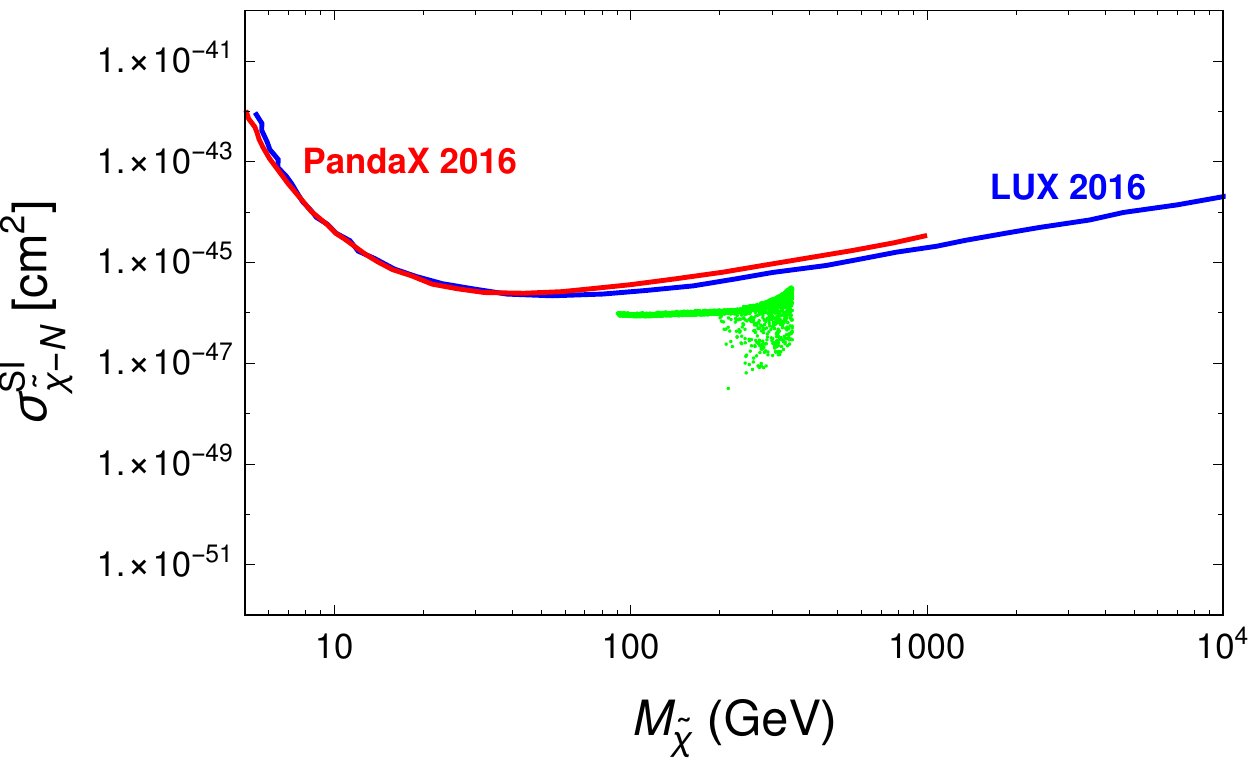}
\includegraphics[scale=0.34]{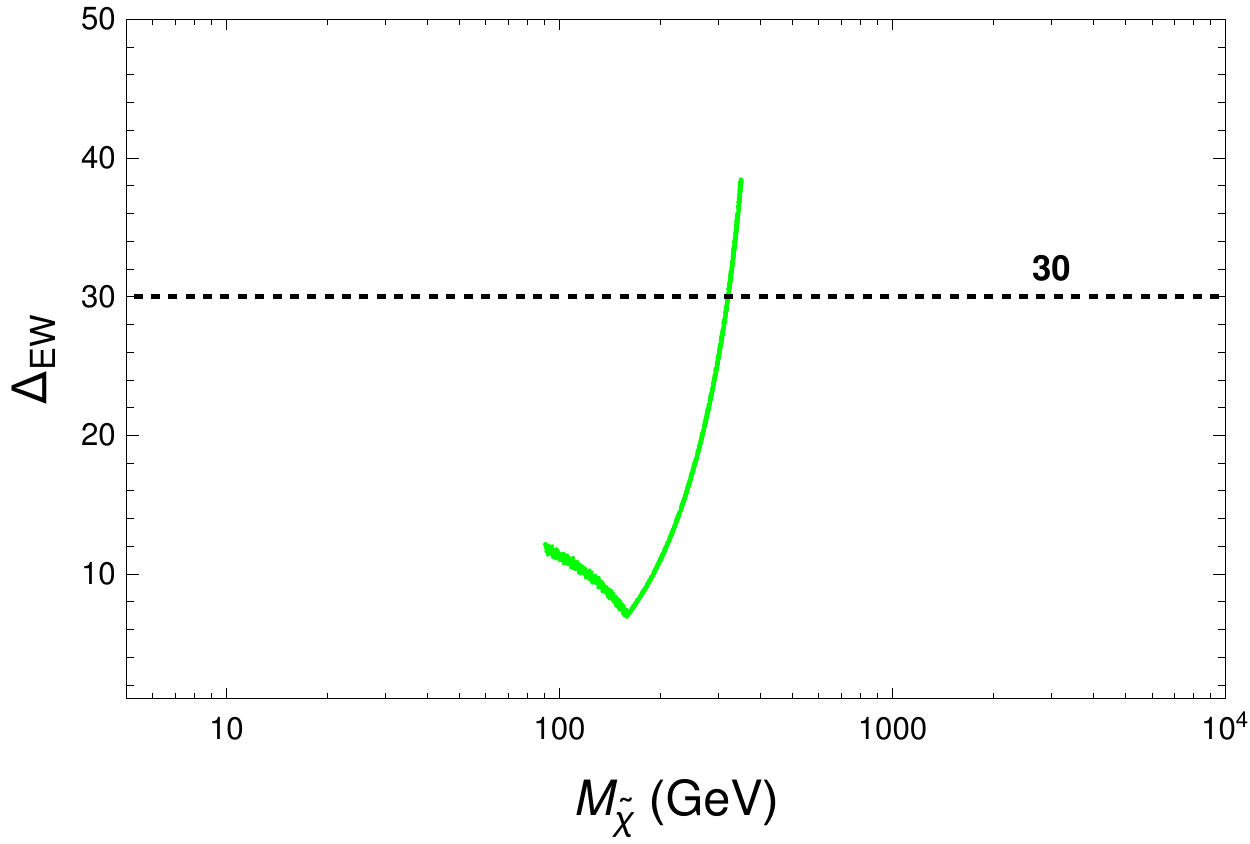}
\includegraphics[scale=0.36]{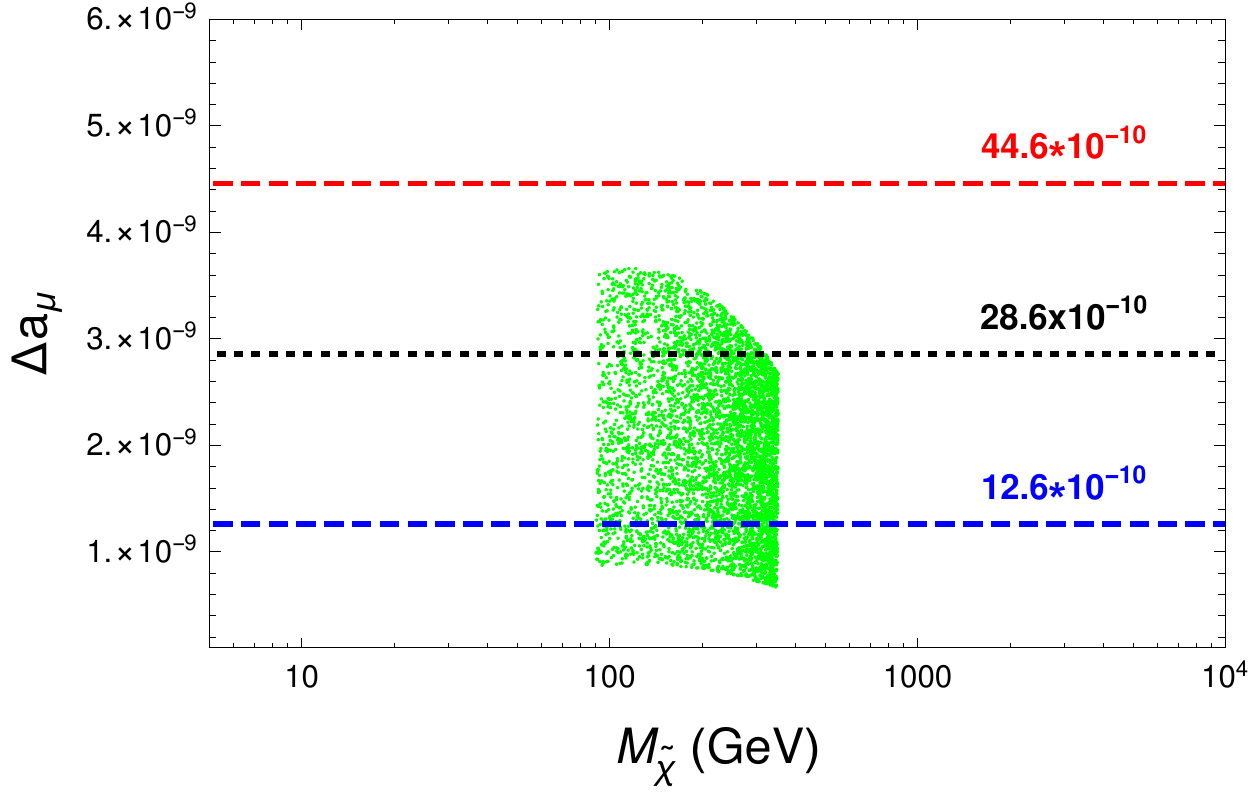}
\end{center}
\caption{The spin-independent elastic dark matter-nucleon scattering cross section, fine-tuning measure,
and muon anomalous magnetic moment versus the LSP neutralino mass for Case D.}
\label{fig:HiggsinoLSP}
\end{figure}

\begin{figure} [H]
\begin{center}
\includegraphics[width=0.3\textwidth]{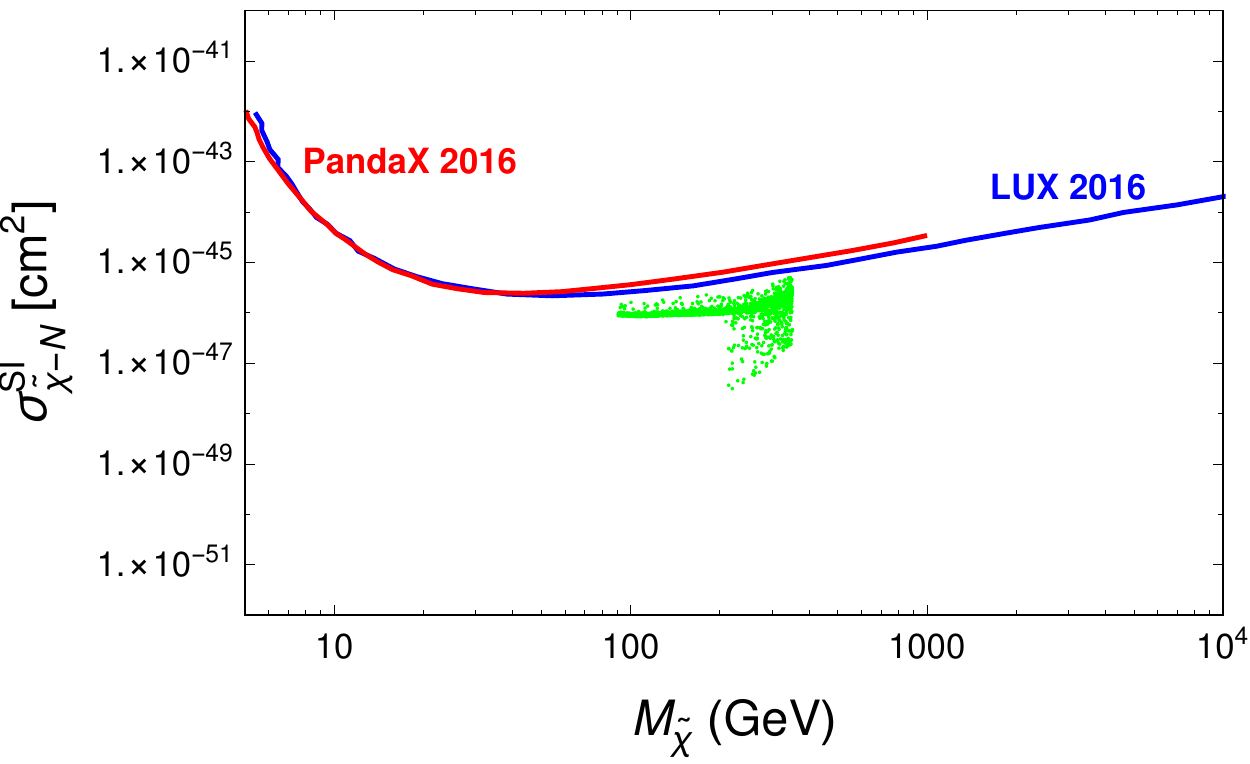}
\includegraphics[scale=0.34]{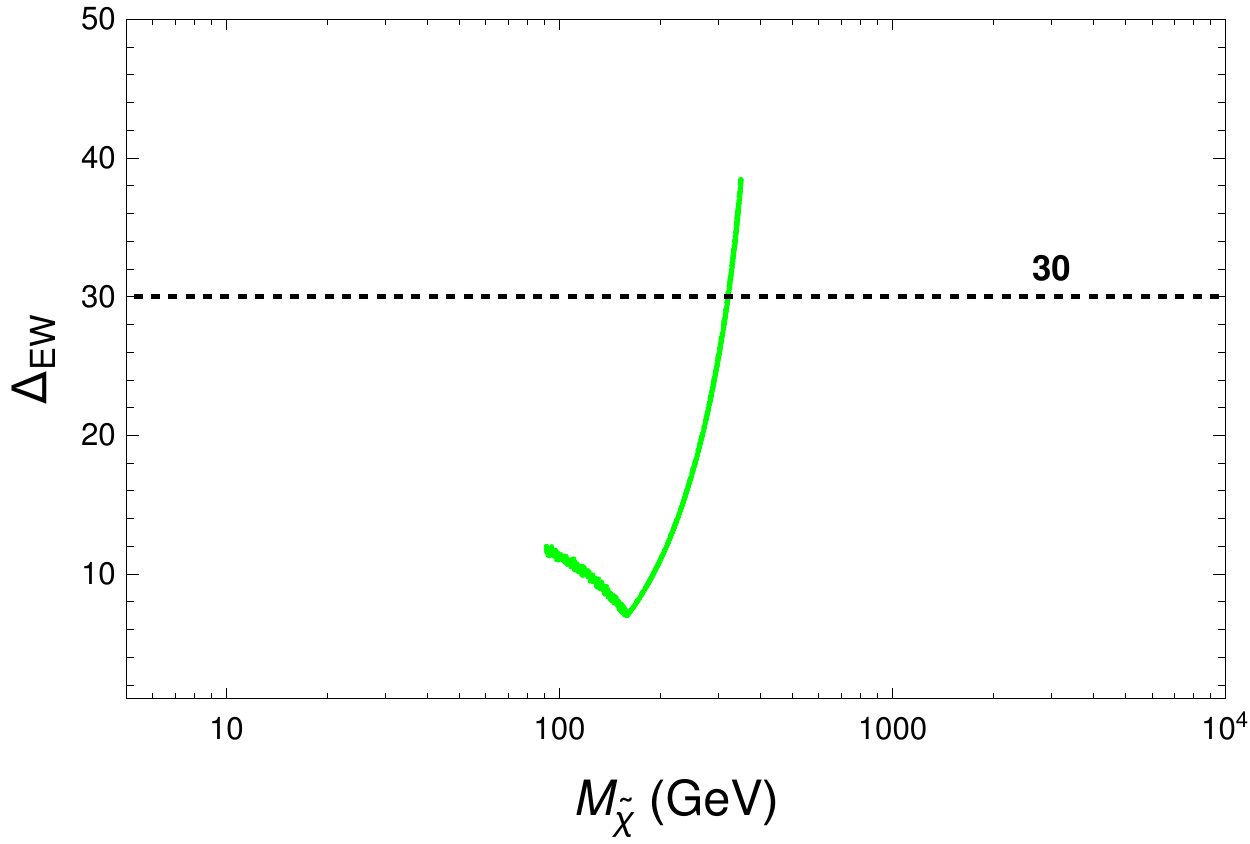}
\includegraphics[scale=0.36]{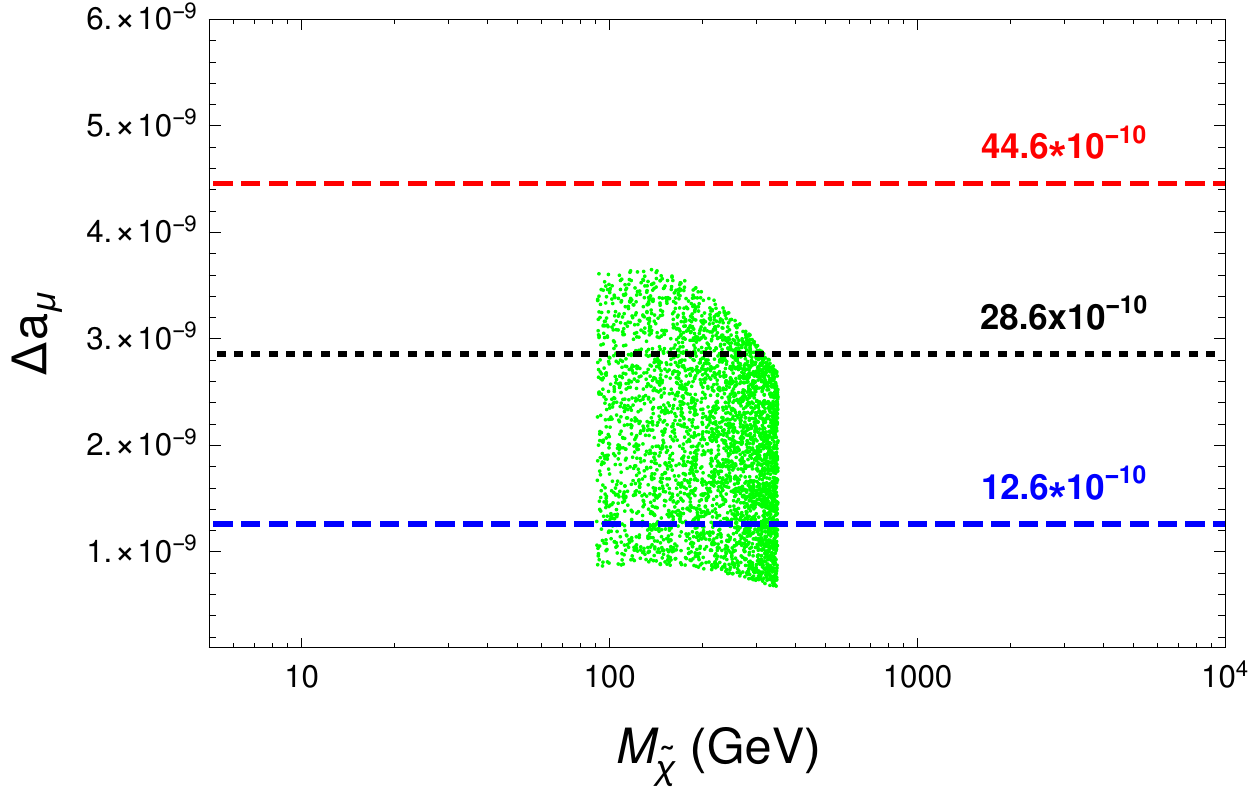}
\end{center}
\caption{The spin-independent elastic dark matter-nucleon scattering cross section, fine-tuning measure,
and muon anomalous magnetic moment versus the LSP neutralino mass for Case E.}
\label{fig:HiggsinoLSP_StauCoan}
\end{figure}

\begin{figure} [H]
\begin{center}
\includegraphics[width=0.3\textwidth]{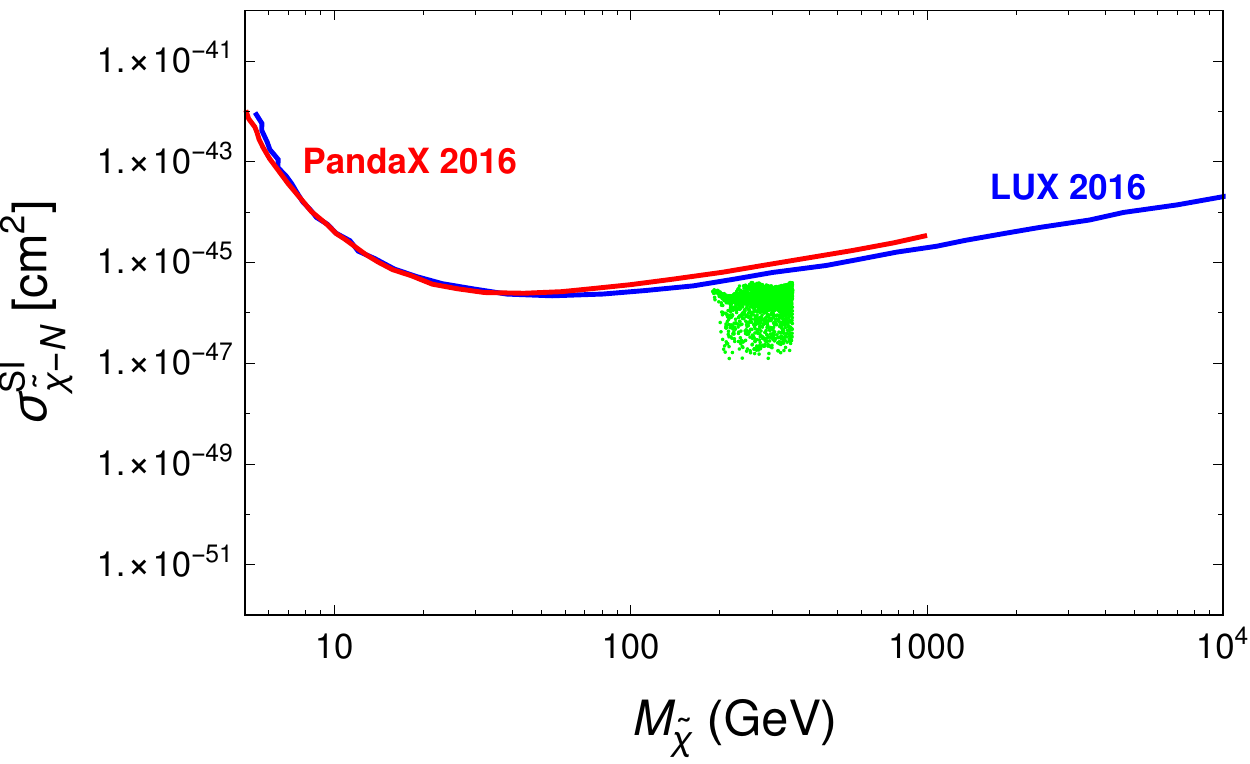}
\includegraphics[scale=0.34]{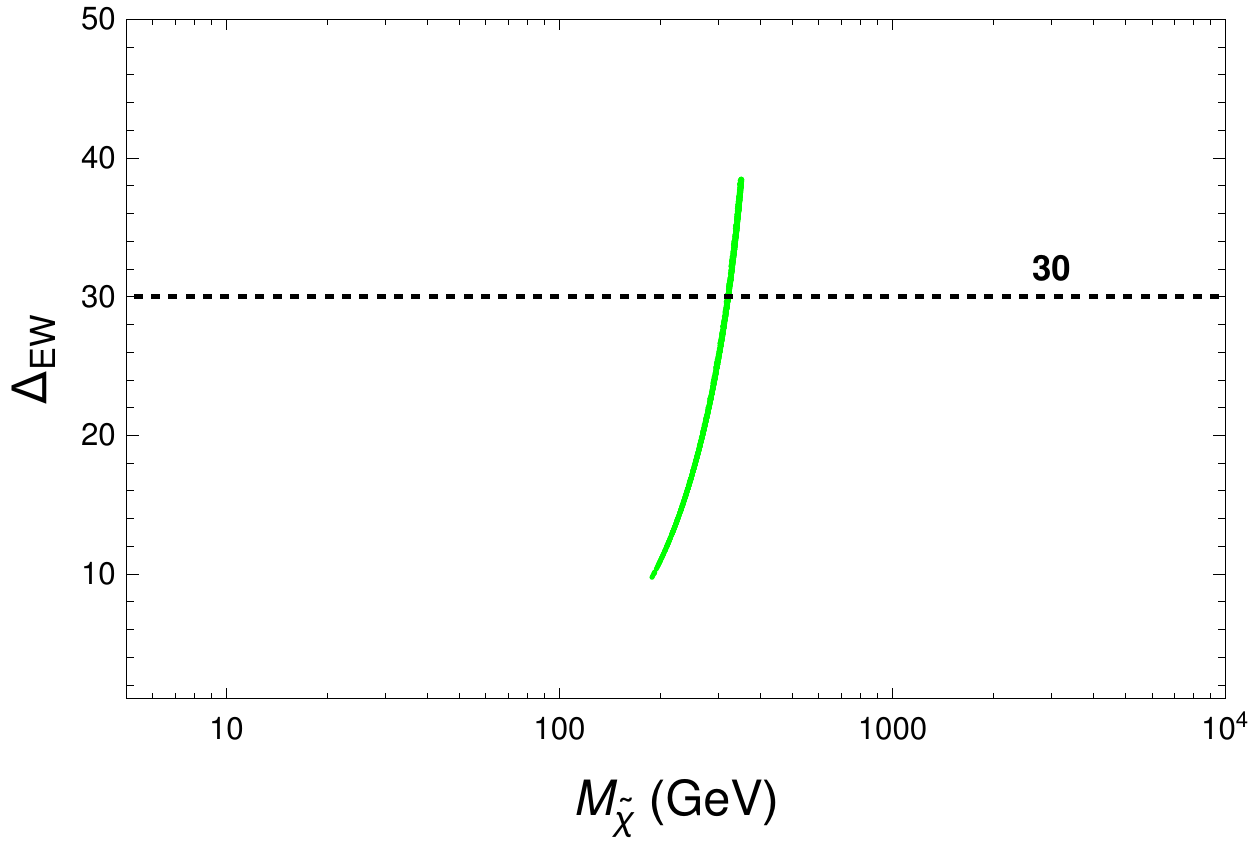}
\includegraphics[scale=0.36]{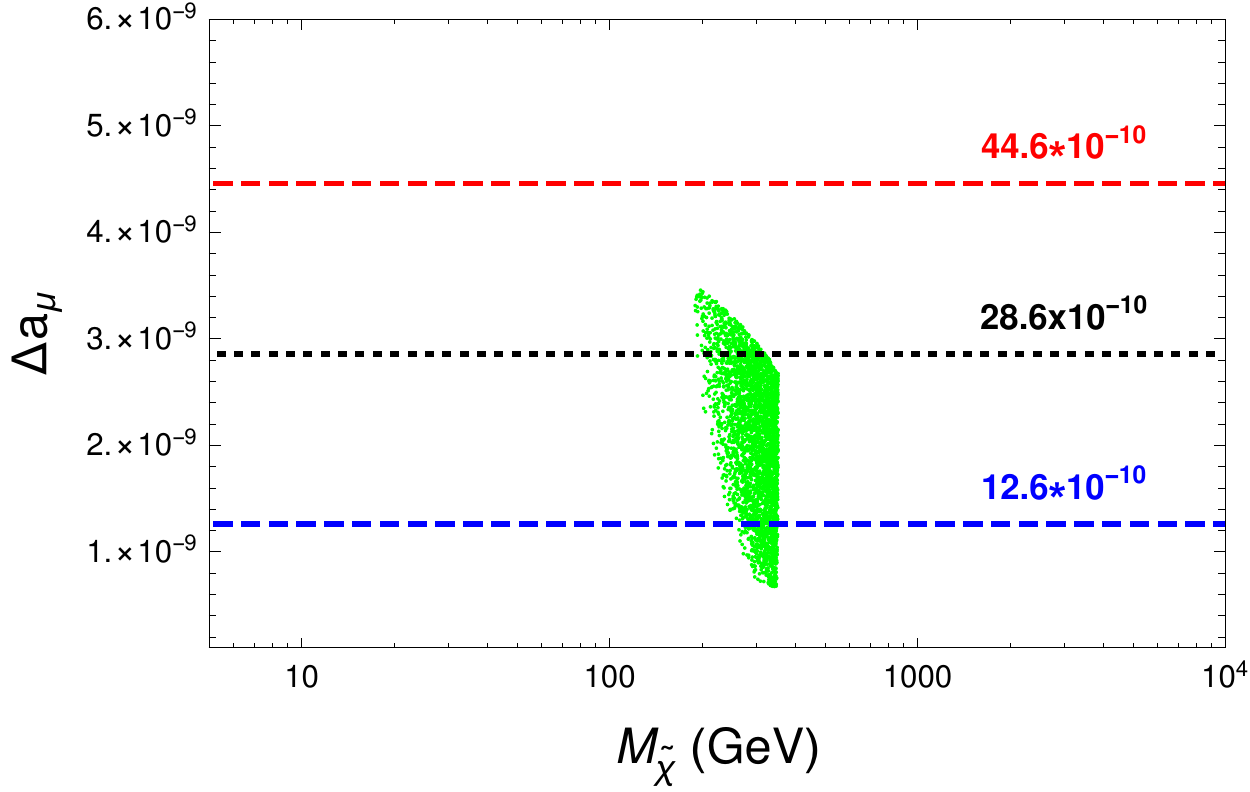}
\end{center}
\caption{The spin-independent elastic dark matter-nucleon scattering cross section, fine-tuning measure,
and muon anomalous magnetic moment versus the LSP neutralino mass for Case F.}
\label{fig:HiggsinoLSP_AReson}
\end{figure}

In our numerical studies, we consider the following six Cases:

\begin{itemize}
	
\item Case A with general scan for the phenomenological preferred parameter space. To explain the anomalous magnetic moment of the muon and have the small low energy fine-tuning measures, we consider the input parameters given in Table \ref{table:GeneralScan}, and present the spin-independent elastic dark matter-nucleon scattering cross section, fine-tuning measure, and muon anomalous magnetic moment versus the LSP neutralino mass in Fig.~\ref{fig:GeneralScan}. Indeed, large parameter space is excluded by the LUX and PANDAX experiments. Interestingly, there are four viable regions: the $Z$ resonance region with $m_{\chi^0_1} \simeq M_Z/2\simeq 45.5~{\rm GeV}$, SM Higgs boson resonance region with $m_{\chi^0_1} \simeq m_h/2 \simeq 62.5~{\rm GeV}$, Higgs funnel with $m_{\chi^0_1} \simeq m_{H/A}/2$, and Higgsino LSP. Becuase the soft masses for stau are taken to be relatively heavy, we do not have the light stau coannihilation region here. To be concrete, we present four benchmark points for these four regions respectively in Tables~\ref{table:BMP_GeneralScan_01}, \ref{table:BMP_GeneralScan_02}, \ref{table:BMP_GeneralScan_03}, and \ref{table:BMP_GeneralScan_04}. The corresponding fine-tuning measures are 11.0, 9.9, 25.2, and 21.1, respectively. Thus these points are natural. Also, the muon anomalous magnetic moments for the benchmark points in Tables \ref{table:BMP_GeneralScan_01} and \ref{table:BMP_GeneralScan_02} are close to the central value, while those for the benchmark points in Tables \ref{table:BMP_GeneralScan_03} and \ref{table:BMP_GeneralScan_04} are within $2\sigma$ range. Moreover, when the LSP neutralino mass increases, the fine-tuning measure decreases and increases for $m_{\widetilde{\chi}_{1}^{0}} < 150~{\rm GeV}$ and $m_{\widetilde{\chi}_{1}^{0}} > 150~{\rm GeV}$, respectively. So the fine-tuning measure has a minimum around $m_{\widetilde{\chi}_{1}^{0}}= 150~{\rm GeV}$. The reason is that for  $m_{\widetilde{\chi}_{1}^{0}} < 150~{\rm GeV}$, $\Delta_{EW}$ is given by the fine-tuning measure of $m_{H_{u}^{2}}$, while $m_{\widetilde{\chi}_{1}^{0}} > 150~{\rm GeV}$,  $\Delta_{EW}$ is given by the fine-tuning measure of $\mu$. This conclusion is valid for the Cases with Higgsino LSP as well.

\item Case B with the LSP neutralino and light stau coannihilation, {\it i.e.}, $m_{\tilde{\tau}_1}\thickapprox m_{\tilde{\chi}_1}$. With the input parameters given in Table \ref{table:StauCoan}, we present the spin-independent elastic dark matter-nucleon scattering cross section, fine-tuning measure, and muon anomalous magnetic moment versus the LSP neutralino mass in Fig.~\ref{fig:StauCoan}. Only small parameter space is excluded by the LUX and PANDAX experiments, the fine-tuning measure is around 38.5 since we choose $\mu=400$~GeV, and the muon anomalous magnetic moments for most of the parameter space is within $2\sigma$ range.  To be concrete, we present a benchmark point in Table~\ref{table:BMP_StauCoan} with $\Delta a_{\mu}$ close to central value.

\item Case C with Higgs funnel, {\it i.e.}, $2m_{\tilde{\chi}_1}\thickapprox m_A$. With the input parameters given in Table \ref{table:AReson}, we present the spin-independent elastic dark matter-nucleon scattering cross section, fine-tuning measure, and muon anomalous magnetic moment versus the LSP neutralino mass in Fig.~\ref{fig:AReson}. Similar to the Case B, only small parameter space is excluded by the LUX and PANDAX experiments, the fine-tuning measure is around 38.5, and the muon anomalous magnetic moments for most of the parameter space is within $2\sigma$ range.  Also, we present a benchmark point in Table~\ref{table:BMP_AReson} with $\Delta a_{\mu}$ close to central value.

\item Case D with Higgsino LSP.  With the input parameters given in Table \ref{table:HiggsinoLSP}, we present the spin-independent elastic dark matter-nucleon scattering cross section, fine-tuning measure, and muon anomalous magnetic moment versus the LSP neutralino mass in Fig.~\ref{fig:HiggsinoLSP}. Because the LSP neutralino relic density is small, the LUX and PANDAX experimental constraints are satisfied after rescale. The low energy fine-tuning measure is similar to Case (A), and the muon anomalous magnetic moment can be explained.  Moreover, we present a benchmark point in Table~\ref{table:BMP_HiggsinoLSP} with fine-tuning measure around 8.87, and $\Delta a_{\mu}$ around central value.

\item Case E is a hybrid scenario with light stau coannihilation and Higgsino LSP. With the input parameters given in Table~\ref{table:HiggsinoLSP_StauCoan}, we present the spin-independent elastic dark matter-nucleon scattering cross section, fine-tuning measure, and muon anomalous magnetic moment versus the LSP neutralino mass in Fig.~\ref{fig:HiggsinoLSP_StauCoan}, which are similar to the Case D. We also present a benchmark point in Table~\ref{table:BMP_HiggsinoLSP_StauCoan} with fine-tuning measure around 9.05, and $\Delta a_{\mu}$ close to central value.

\item Case F is another hybrid scenario with Higgs funnel and Higgsino LSP. With the input parameters given in Table~\ref{table:HiggsinoLSP_AReson}, we present the spin-independent elastic dark matter-nucleon scattering cross section, fine-tuning measure, and muon anomalous magnetic moment versus the LSP neutralino mass in Fig.~\ref{fig:HiggsinoLSP_AReson}. This Case is similar to the Case D except that the LSP neutralino mass is larger than about 180~GeV. We present a benchmark point in Table~\ref{table:BMP_HiggsinoLSP_AReson} with fine-tuning measure around 11.7, and $\Delta a_{\mu}$ close to central value.

\end{itemize}

\begin{table}[H]
	\small
	\centering
	\caption{The input parameters for Case A. }
	\begin{tabular}{|c|c||c|c||c|c||c|c||c|c||c|c||} \hline		
		 $tan\beta$&$[2,60]$&$M_1$&$[50,300]$&$m_{\tilde{Q},\tilde{U},\tilde{D},1\&2}$&$1.2\times10^3$&$m_{\tilde{L},\tilde{E},1\&2}$&$500$\\ \hline
		 $\mu$&$[100,350]$&$M_2$&$400$&$m_{\tilde{Q},3}$&$900$&$m_{\tilde{L},3}$&$500$\\ \hline		
		 $B_{\mu}$&$2\times10^4$&$M_3$&$600$&$m_{\tilde{U},3}$&$900$&$m_{\tilde{E},3}$&$500$\\ \hline
		 $l_{up}$&$[-0.50,-0.01]$&$M_D$&$3\times10^3$&$m_{\tilde{D},3}$&$900$&$m_{\Phi}$&$2\times10^3$\\ \hline
	\end{tabular}
	\label{table:GeneralScan}
\end{table}

\begin{table}[H]
	\small
	\centering
	\caption{The particle spectrum (in GeV) of the benchmark point for Case A with $Z$ resonance. Here, $tan\beta = 40.6533$, $l_{up} = -0.0215602$, $\mu = 129.592$, $M1 =53.4994$, $\Omega_{\chi}h^2 = 0.004233$, $\sigma_{SI} = 2.15906\times10^{-45}~{\rm cm}^2$, $\Delta_{EW} = 10.9909$, $\Delta a_\mu = 3.75112\times10^{-9}$}
	\begin{tabular}{|c|c||c|c||c|c||c|c||c|c||c|c||} \hline		
		 $h$&$124.822$&$\widetilde{\chi}_{1}^{0}$&$46.2694$&$\widetilde{\nu}_{\tau}$&$496.784$&$\widetilde{\tau}_{1}$&$490.723$&$\widetilde{t}_{1}$&$1185.8$&$\widetilde{b}_{1}$&$1179.46$\\\hline
		 $H$&$862.824$&$\widetilde{\chi}_{2}^{0}$&$131.972$&$\widetilde{\nu}_{\mu}$&$499.421$&$\widetilde{\mu}_{1}$&$502.8$&$\widetilde{t}_{2}$&$1193.86$&$\widetilde{b}_{2}$&$1190.84$\\\hline
		 $A$&$875.399$&$\widetilde{\chi}_{3}^{0}$&$142.3$&$\widetilde{\nu}_{e}$&$499.43$&$\widetilde{e}_{1}$&$502.905$&$\widetilde{c}_{1}$&$1453.17$&$\widetilde{s}_{1}$&$1453.51$\\\hline
		 $H^{\pm}$&$875.372$&$\widetilde{\chi}_{4}^{0}$&$429.516$&$O_\phi$&$6095.86$&$\widetilde{e}_{2}$&$506.258$&$\widetilde{u}_{1}$&$1453.17$&$\widetilde{d}_{1}$&$1453.52$\\\hline
		 $\widetilde{g}_1$&$2913.3$&$\widetilde{\chi}_{1}^{\pm}$&$126.274$&$O_\sigma$&$1445.72$&$\widetilde{\mu}_{2}$&$506.335$&$\widetilde{u}_{2}$&$1457.07$&$\widetilde{s}_{2}$&$1459.43$\\\hline
		 $\widetilde{g}_2$&$3449.35$&$\widetilde{\chi}_{2}^{\pm}$&$429.866$&$-$&$-$&$\widetilde{\tau}_{2}$&$510.307$&$\widetilde{c}_{2}$&$1457.07$&$\widetilde{d}_{2}$&$1459.44$\\\hline				 
	\end{tabular}
	\label{table:BMP_GeneralScan_01}
\end{table}

\begin{table}[H]
	\small
	\centering
	\caption{The particle spectrum (in GeV) of the benchmark point for Case A with SM Higgs boson resonance. Here, $tan\beta = 41.5854$, $l_{up} = -0.0261469$, $\mu = 144.785$, $M1 =69.7085$, $\Omega_{\chi}h^2 = 0.0033$, $\sigma_{SI} = 2.36553\times10^{-45}~{\rm cm}^2$, $\Delta_{EW} = 9.89334$, $\Delta a_\mu = 3.82592\times10^{-9}$.}
	\begin{tabular}{|c|c||c|c||c|c||c|c||c|c||c|c||} \hline		
		 $h$&$126.824$&$\widetilde{\chi}_{1}^{0}$&$61.5097$&$\widetilde{\nu}_{\tau}$&$496.675$&$\widetilde{\tau}_{1}$&$489.249$&$\widetilde{t}_{1}$&$1185.8$&$\widetilde{b}_{1}$&$1178.36$\\\hline
		 $H$&$868.333$&$\widetilde{\chi}_{2}^{0}$&$146.838$&$\widetilde{\nu}_{\mu}$&$499.424$&$\widetilde{\mu}_{1}$&$502.791$&$\widetilde{t}_{2}$&$1193.69$&$\widetilde{b}_{2}$&$1191.36$\\\hline
		 $A$&$881.383$&$\widetilde{\chi}_{3}^{0}$&$156.821$&$\widetilde{\nu}_{e}$&$499.434$&$\widetilde{e}_{1}$&$502.924$&$\widetilde{c}_{1}$&$1453.17$&$\widetilde{s}_{1}$&$1453.51$\\\hline
		 $H^{\pm}$&$882.138$&$\widetilde{\chi}_{4}^{0}$&$429.943$&$O_\phi$&$6095.86$&$\widetilde{e}_{2}$&$506.26$&$\widetilde{u}_{1}$&$1453.17$&$\widetilde{d}_{1}$&$1453.52$\\\hline
		 $\widetilde{g}_1$&$2913.3$&$\widetilde{\chi}_{1}^{\pm}$&$140.885$&$O_\sigma$&$1445.72$&$\widetilde{\mu}_{2}$&$506.362$&$\widetilde{u}_{2}$&$1457.07$&$\widetilde{s}_{2}$&$1459.43$\\\hline
		 $\widetilde{g}_2$&$3449.35$&$\widetilde{\chi}_{2}^{\pm}$&$430.288$&$-$&$-$&$\widetilde{\tau}_{2}$&$511.403$&$\widetilde{c}_{2}$&$1457.07$&$\widetilde{d}_{2}$&$1459.44$\\\hline				 
	\end{tabular}
	\label{table:BMP_GeneralScan_02}
\end{table}

\begin{table}[H]
	\small
	\centering
	\caption{The particle spectrum (in GeV) of the benchmark point for Case A with Higgs funnel. Here, $\tan\beta = 17.3441$, $l_{up} = -0.0264043$, $\mu = 323.513$, $M_1 =295.33$, $\Omega_{\chi}h^2 = 0.000161$, $\sigma_{SI} = 1.39489\times10^{-44}~{\rm cm}^2$, $\Delta_{EW} = 25.1734$, and $\Delta a_\mu = 1.30454\times10^{-9}$.}
	\begin{tabular}{|c|c||c|c||c|c||c|c||c|c||c|c||} \hline		
		 $h$&$126.325$&$\widetilde{\chi}_{1}^{0}$&$267.175$&$\widetilde{\nu}_{\tau}$&$499.367$&$\widetilde{\tau}_{1}$&$494.264$&$\widetilde{t}_{1}$&$1186.81$&$\widetilde{b}_{1}$&$1184.18$\\\hline
		 $H$&$546.53$&$\widetilde{\chi}_{2}^{0}$&$321.976$&$\widetilde{\nu}_{\mu}$&$499.74$&$\widetilde{\mu}_{1}$&$503.339$&$\widetilde{t}_{2}$&$1196.45$&$\widetilde{b}_{2}$&$1196.13$\\\hline
		 $A$&$549.317$&$\widetilde{\chi}_{3}^{0}$&$332.469$&$\widetilde{\nu}_{e}$&$499.741$&$\widetilde{e}_{1}$&$503.449$&$\widetilde{c}_{1}$&$1453.24$&$\widetilde{s}_{1}$&$1453.49$\\\hline
		 $H^{\pm}$&$563.304$&$\widetilde{\chi}_{4}^{0}$&$446.96$&$O_\phi$&$6095.86$&$\widetilde{e}_{2}$&$506.529$&$\widetilde{u}_{1}$&$1453.24$&$\widetilde{d}_{1}$&$1453.49$\\\hline
		 $\widetilde{g}_1$&$2913.3$&$\widetilde{\chi}_{1}^{\pm}$&$300.721$&$O_\sigma$&$1445.72$&$\widetilde{\mu}_{2}$&$506.634$&$\widetilde{u}_{2}$&$1457.07$&$\widetilde{s}_{2}$&$1459.41$\\\hline
		 $\widetilde{g}_2$&$3449.35$&$\widetilde{\chi}_{2}^{\pm}$&$446.386$&$-$&$-$&$\widetilde{\tau}_{2}$&$514.393$&$\widetilde{c}_{2}$&$1457.07$&$\widetilde{d}_{2}$&$1459.42$\\\hline				 
	\end{tabular}
	\label{table:BMP_GeneralScan_03}
\end{table}

\begin{table}[H]
	\small
	\centering
	\caption{The particle spectrum (in GeV) of the benchmark point for Case A with Higgsino LSP.
		Here, $\tan\beta = 16.5305$, $l_{up} = -0.0272574$, $\mu = 296.024$, $M_1 =299.181$, $\Omega_{\chi}h^2 = 0.00065$, $\sigma_{SI} = 1.78553\times10^{-44}~{\rm cm}^2$, $\Delta_{EW} = 21.0772$, and $\Delta a_\mu = 1.28535\times10^{-9}$.}
	\begin{tabular}{|c|c||c|c||c|c||c|c||c|c||c|c||} \hline		
		 $h$&$126.694$&$\widetilde{\chi}_{1}^{0}$&$255.019$&$\widetilde{\nu}_{\tau}$&$499.414$&$\widetilde{\tau}_{1}$&$495.585$&$\widetilde{t}_{1}$&$1186.74$&$\widetilde{b}_{1}$&$1185.03$\\\hline
		 $H$&$536.578$&$\widetilde{\chi}_{2}^{0}$&$305.146$&$\widetilde{\nu}_{\mu}$&$499.76$&$\widetilde{\mu}_{1}$&$503.371$&$\widetilde{t}_{2}$&$1196.5$&$\widetilde{b}_{2}$&$1195.46$\\\hline
		 $A$&$539.181$&$\widetilde{\chi}_{3}^{0}$&$315.396$&$\widetilde{\nu}_{e}$&$499.761$&$\widetilde{e}_{1}$&$503.456$&$\widetilde{c}_{1}$&$1453.25$&$\widetilde{s}_{1}$&$1453.49$\\\hline
		 $H^{\pm}$&$552.227$&$\widetilde{\chi}_{4}^{0}$&$442.047$&$O_\phi$&$6095.86$&$\widetilde{e}_{2}$&$506.542$&$\widetilde{u}_{1}$&$1453.25$&$\widetilde{d}_{1}$&$1453.49$\\\hline
		 $\widetilde{g}_1$&$2913.3$&$\widetilde{\chi}_{1}^{\pm}$&$278.208$&$O_\sigma$&$1445.72$&$\widetilde{\mu}_{2}$&$506.622$&$\widetilde{u}_{2}$&$1457.07$&$\widetilde{s}_{2}$&$1459.41$\\\hline
		 $\widetilde{g}_2$&$3449.35$&$\widetilde{\chi}_{2}^{\pm}$&$441.565$&$-$&$-$&$\widetilde{\tau}_{2}$&$513.222$&$\widetilde{c}_{2}$&$1457.07$&$\widetilde{d}_{2}$&$1459.42$\\\hline				 
	\end{tabular}
	\label{table:BMP_GeneralScan_04}
\end{table}

\begin{table}[H]
	\small
	\centering
	\caption{The input parameters for Case B with $M_{X} \subset [M_1-10,~M_1+50]$.}
	\begin{tabular}{|c|c||c|c||c|c||c|c||c|c||c|c||} \hline		
		 $\tan\beta$&$[10,40]$&$M_1$&$[100,400]$&$m_{\tilde{Q},\tilde{U},\tilde{D},1\&2}$&$\sqrt{2}\times10^3$&$m_{\tilde{L},\tilde{E},1\&2}$&$500$\\ \hline
		 $\mu$&$400$&$M_2$&$400$&$m_{\tilde{Q},3}$&$\sqrt{1.2}\times10^3$&$m_{\tilde{L},3}$&$500$\\ \hline		
		 $B_{\mu}$&$2\times10^4$&$M_3$&$600$&$m_{\tilde{U},3}$&$\sqrt{1.2}\times10^3$&$m_{\tilde{E},3}$&$M_{X}$\\ \hline
		 $l_{up}$&$[-0.04,-0.01]$&$M_D$&$3\times10^3$&$m_{\tilde{D},3}$&$\sqrt{1.2}\times10^3$&$m_{\Phi}$&$2\times10^3$\\ \hline
	\end{tabular}
	\label{table:StauCoan}
\end{table}

\begin{table}[H]
	\small
	\centering
	\caption{The particle spectrum (in GeV) of the benchmark point for Case B.
		Here, $\tan\beta = 37.8478$, $l_{up} = -0.0363711$, $M_1 = 147.266$, $m_{\tilde{E},3} = 150.701$, $\Omega_{\chi}h^2 = 0.028686$, $\sigma_{SI} = 2.92858\times10^{-46}~{\rm cm}^2$, $\Delta_{EW} = 38.4838$, and $\Delta a_\mu = 2.61376\times10^{-9}$.}

	\begin{tabular}{|c|c||c|c||c|c||c|c||c|c||c|c||} \hline		
		 $h$&$125.888$&$\widetilde{\chi}_{1}^{0}$&$143.522$&$\widetilde{\nu}_{\tau}$&$497.721$&$\widetilde{\tau}_{1}$&$144.616$&$\widetilde{t}_{1}$&$1354.76$&$\widetilde{b}_{1}$&$1341.03$\\ \hline
		 $H$&$787.832$&$\widetilde{\chi}_{2}^{0}$&$355.474$&$\widetilde{\nu}_{\mu}$&$499.26$&$\widetilde{\mu}_{1}$&$502.69$&$\widetilde{t}_{2}$&$1362.57$&$\widetilde{b}_{2}$&$1368.96$\\ \hline
		 $A$&$798.515$&$\widetilde{\chi}_{3}^{0}$&$409.602$&$\widetilde{\nu}_{e}$&$499.267$&$\widetilde{e}_{1}$&$503.436$&$\widetilde{c}_{1}$&$1648.91$&$\widetilde{s}_{1}$&$1649.14$\\ \hline
		 $H^{\pm}$&$813.684$&$\widetilde{\chi}_{4}^{0}$&$470.665$&$O_\phi$&$6082.4$&$\widetilde{e}_{2}$&$506.079$&$\widetilde{u}_{1}$&$1648.91$&$\widetilde{d}_{1}$&$1649.15$\\ \hline
		 $\widetilde{g}_1$&$2911.4$&$\widetilde{\chi}_{1}^{\pm}$&$355.447$&$O_\sigma$&$1445.72$&$\widetilde{\mu}_{2}$&$506.799$&$\widetilde{u}_{2}$&$1653.5$&$\widetilde{d}_{2}$&$1655.56$\\ \hline
		 $\widetilde{g}_2$&$3443.67$&$\widetilde{\chi}_{2}^{\pm}$&$471.278$&$-$&$-$&$\widetilde{\tau}_{2}$&$507.561$&$\widetilde{c}_{2}$&$1653.5$&$\widetilde{s}_{2}$&$1655.56$\\ \hline				
	\end{tabular}
	\label{table:BMP_StauCoan}
\end{table}

\begin{table}[H]
	\small
	\centering
	\caption{The input parameters for Case C with $M^2_{X} \subset \left[(0.34M_1)^2, (0.90M_1)^2\right]$.}
	\begin{tabular}{|c|c||c|c||c|c||c|c||c|c||c|c||} \hline		
		 $\tan\beta$&$[10,40]$&$M_1$&$[200,400]$&$m_{\tilde{Q},\tilde{U},\tilde{D},1\&2}$&$\sqrt{2}\times10^3$&$m_{\tilde{L},\tilde{E},1\&2}$&$500$\\ \hline
		 $\mu$&$400$&$M_2$&$400$&$m_{\tilde{Q},3}$&$\sqrt{1.2}\times10^3$&$m_{\tilde{L},3}$&$500$\\ \hline		
		 $B_{\mu}$&$M_X^2$&$M_3$&$600$&$m_{\tilde{U},3}$&$\sqrt{1.2}\times10^3$&$m_{\tilde{E},3}$&$400$\\ \hline
		 $l_{up}$&$[-0.04,-0.01]$&$M_D$&$3\times10^3$&$m_{\tilde{D},3}$&$\sqrt{1.2}\times10^3$&$m_{\Phi}$&$2\times10^3$\\ \hline
	\end{tabular}
	\label{table:AReson}
\end{table}

\begin{table}[H]
	\small
	\centering
	\caption{The particle spectrum (in GeV) of the benchmark point for Case C.
		Here, $\tan\beta = 37.4813$, $l_{up} = -0.022275$, $B_\mu = 13747.8$, $M_1 =318.273$, $\Omega_{\chi}h^2 = 0.000619$, $\sigma_{SI} = 7.94557\times10^{-45}~{\rm cm}^2$, $\Delta_{EW} = 38.4838$, and $\Delta a_\mu = 2.58449\times10^{-9}$.}
	\begin{tabular}{|c|c||c|c||c|c||c|c||c|c||c|c||} \hline		
		 $h$&$125.693$&$\widetilde{\chi}_{1}^{0}$&$303.924$&$\widetilde{\nu}_{\tau}$&$497.907$&$\widetilde{\tau}_{1}$&$392.666$&$\widetilde{t}_{1}$&$1354.76$&$\widetilde{b}_{1}$&$1341.29$\\\hline
		 $H$&$617.195$&$\widetilde{\chi}_{2}^{0}$&$362.531$&$\widetilde{\nu}_{\mu}$&$499.626$&$\widetilde{\mu}_{1}$&$503.084$&$\widetilde{t}_{2}$&$1362.62$&$\widetilde{b}_{2}$&$1368.99$\\\hline
		 $A$&$630.019$&$\widetilde{\chi}_{3}^{0}$&$409.204$&$\widetilde{\nu}_{e}$&$499.632$&$\widetilde{e}_{1}$&$503.818$&$\widetilde{c}_{1}$&$1648.99$&$\widetilde{s}_{1}$&$1649.11$\\\hline
		 $H^{\pm}$&$649.769$&$\widetilde{\chi}_{4}^{0}$&$472.22$&$O_\phi$&$6082.4$&$\widetilde{e}_{2}$&$506.449$&$\widetilde{u}_{1}$&$1648.99$&$\widetilde{d}_{1}$&$1649.12$\\\hline
		 $\widetilde{g}_1$&$2911.4$&$\widetilde{\chi}_{1}^{\pm}$&$355.394$&$O_\sigma$&$1445.72$&$\widetilde{\mu}_{2}$&$507.16$&$\widetilde{u}_{2}$&$1653.48$&$\widetilde{d}_{2}$&$1655.54$\\\hline
		 $\widetilde{g}_2$&$3443.67$&$\widetilde{\chi}_{2}^{\pm}$&$471.153$&$-$&$-$&$\widetilde{\tau}_{2}$&$511.568$&$\widetilde{c}_{2}$&$1653.48$&$\widetilde{s}_{2}$&$1655.54$\\\hline				 
	\end{tabular}
	\label{table:BMP_AReson}
\end{table}

\begin{table}[H]
	\small
	\centering
	\caption{The input parameters for Case D. }
	\begin{tabular}{|c|c||c|c||c|c||c|c||c|c||c|c||} \hline		
		 $\tan\beta$&$[10,40]$&$M_1$&$1000$&$m_{\tilde{Q},\tilde{U},\tilde{D},1\&2}$&$\sqrt{2}\times10^3$&$m_{\tilde{L},\tilde{E},1\&2}$&$500$\\ \hline
		 $\mu$&$[100,400]$&$M_2$&$400$&$m_{\tilde{Q},3}$&$\sqrt{1.2}\times10^3$&$m_{\tilde{L},3}$&$500$\\ \hline		
		 $B_{\mu}$&$2\times10^4$&$M_3$&$600$&$m_{\tilde{U},3}$&$\sqrt{1.2}\times10^3$&$m_{\tilde{E},3}$&$500$\\ \hline
		 $l_{up}$&$[-0.04,-0.01]$&$M_D$&$3\times10^3$&$m_{\tilde{D},3}$&$\sqrt{1.2}\times10^3$&$m_{\Phi}$&$2\times10^3$\\ \hline
	\end{tabular}
	\label{table:HiggsinoLSP}
\end{table}

\begin{table}[H]
	\small
	\centering
	\caption{The particle spectrum (in GeV) of the benchmark point for Case D.
		Here, $\tan\beta = 32.2887$, $l_{up} = -0.0206558$, $\mu = 192.031$, $\Omega_{\chi}h^2 = 0.002068$, $\sigma_{SI} = 5.59253\times10^{-45}~{\rm cm}^2$, $\Delta_{EW} = 8.86957$, and $\Delta a_\mu = 2.84023\times10^{-9}$.}
	\begin{tabular}{|c|c||c|c||c|c||c|c||c|c||c|c||} \hline		
		 $h$&$125.119$&$\widetilde{\chi}_{1}^{0}$&$180.688$&$\widetilde{\nu}_{\tau}$&$498.917$&$\widetilde{\tau}_{1}$&$493.798$&$\widetilde{t}_{1}$&$1354.95$&$\widetilde{b}_{1}$&$1350.75$\\\hline
		 $H$&$759.686$&$\widetilde{\chi}_{2}^{0}$&$201.64$&$\widetilde{\nu}_{\mu}$&$500.506$&$\widetilde{\mu}_{1}$&$506.618$&$\widetilde{t}_{2}$&$1363.46$&$\widetilde{b}_{2}$&$1362.21$\\\hline
		 $A$&$768.028$&$\widetilde{\chi}_{3}^{0}$&$433.02$&$\widetilde{\nu}_{e}$&$500.512$&$\widetilde{e}_{1}$&$507.172$&$\widetilde{c}_{1}$&$1649.2$&$\widetilde{s}_{1}$&$1649.16$\\\hline
		 $H^{\pm}$&$772.883$&$\widetilde{\chi}_{4}^{0}$&$990.236$&$O_\phi$&$6082.4$&$\widetilde{e}_{2}$&$507.383$&$\widetilde{u}_{1}$&$1649.2$&$\widetilde{d}_{1}$&$1649.16$\\\hline
		 $\widetilde{g}_1$&$2911.4$&$\widetilde{\chi}_{1}^{\pm}$&$186.564$&$O_\sigma$&$1445.72$&$\widetilde{\mu}_{2}$&$507.92$&$\widetilde{c}_{2}$&$1653.49$&$\widetilde{s}_{2}$&$1655.54$\\\hline
		 $\widetilde{g}_2$&$3443.67$&$\widetilde{\chi}_{2}^{\pm}$&$433.843$&$-$&$-$&$\widetilde{\tau}_{2}$&$515.737$&$\widetilde{u}_{2}$&$1653.49$&$\widetilde{d}_{2}$&$1655.55$\\\hline				 
	\end{tabular}
	\label{table:BMP_HiggsinoLSP}
\end{table}

\begin{table}[H]
	\small
	\centering
	\caption{The input parameters for Case E with $M_{X} \subset [\mu,~\mu+100]$.}
	\begin{tabular}{|c|c||c|c||c|c||c|c||c|c||c|c||} \hline		
		 $\tan\beta$&$[10,40]$&$M_1$&$1000$&$m_{\tilde{Q},\tilde{U},\tilde{D},1\&2}$&$\sqrt{2}\times10^3$&$m_{\tilde{L},\tilde{E},1\&2}$&$500$\\ \hline
		 $\mu$&$[100,400]$&$M_2$&$400$&$m_{\tilde{Q},3}$&$\sqrt{1.2}\times10^3$&$m_{\tilde{L},3}$&$500$\\ \hline		
		 $B_{\mu}$&$2\times10^4$&$M_3$&$600$&$m_{\tilde{U},3}$&$\sqrt{1.2}\times10^3$&$m_{\tilde{E},3}$&$M_X$\\ \hline
		 $l_{up}$&$[-0.04,-0.01]$&$M_D$&$3\times10^3$&$m_{\tilde{D},3}$&$\sqrt{1.2}\times10^3$&$m_{\Phi}$&$2\times10^3$\\ \hline
	\end{tabular}
	\label{table:HiggsinoLSP_StauCoan}
\end{table}

\begin{table}[H]
	\small
	\centering
	\caption{The particle spectrum (in GeV) of the benchmark point for Case E.
		Here, $\tan\beta = 35.1067$, $l_{up} = -0.0362218$, $\mu = 138.992$, $m_{\tilde{E},3} = 125.942$, $\Omega_{\chi}h^2 = 0.003076$, $\sigma_{SI} = 3.8407\times10^{-45}~{\rm cm}^2$, $\Delta_{EW} = 9.04788$, and $\Delta a_\mu = 3.2067\times10^{-9}$.}
	\begin{tabular}{|c|c||c|c||c|c||c|c||c|c||c|c||} \hline		
		 $h$&$125.354$&$\widetilde{\chi}_{1}^{0}$&$129.548$&$\widetilde{\nu}_{\tau}$&$498.762$&$\widetilde{\tau}_{1}$&$139.268$&$\widetilde{t}_{1}$&$1354.72$&$\widetilde{b}_{1}$&$1351.32$\\\hline
		 $H$&$801.917$&$\widetilde{\chi}_{2}^{0}$&$148.352$&$\widetilde{\nu}_{\mu}$&$500.295$&$\widetilde{\tau}_{2}$&$505.946$&$\widetilde{t}_{2}$&$1363.13$&$\widetilde{b}_{2}$&$1360.43$\\\hline
		 $A$&$811.285$&$\widetilde{\chi}_{3}^{0}$&$431.17$&$\widetilde{\nu}_{e}$&$500.301$&$\widetilde{\mu}_{1}$&$506.826$&$\widetilde{c}_{1}$&$1649.11$&$\widetilde{s}_{1}$&$1649.2$\\\hline
		 $H^{\pm}$&$812.953$&$\widetilde{\chi}_{4}^{0}$&$990.563$&$O_\phi$&$6082.4$&$\widetilde{e}_{1}$&$507.169$&$\widetilde{u}_{1}$&$1649.11$&$\widetilde{d}_{1}$&$1649.2$\\\hline
		 $\widetilde{g}_1$&$2911.4$&$\widetilde{\chi}_{1}^{\pm}$&$135.854$&$O_\sigma$&$1445.72$&$\widetilde{e}_{2}$&$507.598$&$\widetilde{u}_{2}$&$1653.51$&$\widetilde{s}_{2}$&$1655.56$\\\hline
		 $\widetilde{g}_2$&$3443.67$&$\widetilde{\chi}_{2}^{\pm}$&$431.881$&$-$&$-$&$\widetilde{\mu}_{2}$&$507.921$&$\widetilde{c}_{2}$&$1653.51$&$\widetilde{d}_{2}$&$1655.57$\\\hline				 
	\end{tabular}
	\label{table:BMP_HiggsinoLSP_StauCoan}
\end{table}

\begin{table}[H]
	\small
	\centering
	\caption{The input parameters for Case F with $M^2_{X} \subset \left[(0.30\mu)^2, (0.42\mu)^2\right]$. }
	\begin{tabular}{|c|c||c|c||c|c||c|c||c|c||c|c||} \hline		
		 $\tan\beta$&$[10,40]$&$M_1$&$1000$&$m_{\tilde{Q},\tilde{U},\tilde{D},1\&2}$&$\sqrt{2}\times10^3$&$m_{\tilde{L},\tilde{E},1\&2}$&$500$\\ \hline
		 $\mu$&$[100,400]$&$M_2$&$400$&$m_{\tilde{Q},3}$&$\sqrt{1.2}\times10^3$&$m_{\tilde{L},3}$&$500$\\ \hline		
		 $B_{\mu}$&$M_X^2$&$M_3$&$600$&$m_{\tilde{U},3}$&$\sqrt{1.2}\times10^3$&$m_{\tilde{E},3}$&$500$\\ \hline
		 $l_{up}$&$[-0.04,-0.01]$&$M_D$&$3\times10^3$&$m_{\tilde{D},3}$&$\sqrt{1.2}\times10^3$&$m_{\Phi}$&$2\times10^3$\\ \hline
	\end{tabular}
	\label{table:HiggsinoLSP_AReson}
\end{table}

\begin{table}[H]
	\small
	\centering
	\caption{The particle spectrum (in GeV) of the benchmark point for Case F.
		Here, $\tan\beta = 31.6269$, $l_{up} = -0.0232174$, $\mu = 220.941$, $B_\mu = 7450.75$, $\Omega_{\chi}h^2 = 0.000136$, $\sigma_{SI} = 1.92848\times10^{-44}~{\rm cm}^2$, $\Delta_{EW} = 11.7412$, and $\Delta a_\mu = 2.69287\times10^{-9}$.}
	\begin{tabular}{|c|c||c|c||c|c||c|c||c|c||c|c||} \hline		
		 $h$&$126.132$&$\widetilde{\chi}_{1}^{0}$&$207.818$&$\widetilde{\nu}_{\tau}$&$499.38$&$\widetilde{\tau}_{1}$&$492.795$&$\widetilde{t}_{1}$&$1354.97$&$\widetilde{b}_{1}$&$1350.36$\\\hline
		 $H$&$405.181$&$\widetilde{\chi}_{2}^{0}$&$230.439$&$\widetilde{\nu}_{\mu}$&$500.65$&$\widetilde{\mu}_{1}$&$506.409$&$\widetilde{t}_{2}$&$1363.61$&$\widetilde{b}_{2}$&$1363.27$\\\hline
		 $A$&$419.464$&$\widetilde{\chi}_{3}^{0}$&$433.985$&$\widetilde{\nu}_{e}$&$500.654$&$\widetilde{e}_{1}$&$506.893$&$\widetilde{c}_{1}$&$1649.26$&$\widetilde{s}_{1}$&$1649.13$\\\hline
		 $H^{\pm}$&$431.247$&$\widetilde{\chi}_{4}^{0}$&$989.92$&$O_\phi$&$6082.4$&$\widetilde{e}_{2}$&$507.518$&$\widetilde{u}_{1}$&$1649.26$&$\widetilde{d}_{1}$&$1649.13$\\\hline
		 $\widetilde{g}_1$&$2911.4$&$\widetilde{\chi}_{1}^{\pm}$&$213.448$&$O_\sigma$&$1445.72$&$\widetilde{\mu}_{2}$&$507.987$&$\widetilde{u}_{2}$&$1653.47$&$\widetilde{s}_{2}$&$1655.53$\\\hline
		 $\widetilde{g}_2$&$3443.67$&$\widetilde{\chi}_{2}^{\pm}$&$434.901$&$-$&$-$&$\widetilde{\tau}_{2}$&$517.49$&$\widetilde{c}_{2}$&$1653.47$&$\widetilde{d}_{2}$&$1655.53$\\\hline				 
	\end{tabular}
	\label{table:BMP_HiggsinoLSP_AReson}
\end{table}


\section{Conclusions}

We studied the naturalness, dark matter, and muon anomalous magnetic moment in the PDGSSMs. In order to obtain the correct dark matter density and explain the muon anomalous magnetic moment, we found that the low energy fine-tuning measures are larger than about 30 due to strong constraints from the LUX and PANDAX experiments.
Thus, to explore the natural PDGSSMs, we considered multi-component dark matter and then the relic density of the LSP neutralino is smaller than the observed value.
We classified the dark matter models into six kinds:
(i) Case A is a general case, which has small low energy fine-tuning measure and can explain the anomalous magnetic moment of the muon;
(ii) Case B with the LSP neutralino and light stau coannihilation;
(iii) Case C with Higgs funnel;
(iv) Case D with Higgsino LSP;
(v) Case E with light stau coannihilation and Higgsino LSP;
(vi) Case F with Higgs funnel and Higgsino LSP.
We studied these Cases in details, and showed that our models can be natural and consistent with the LUX and PANDAX experiments, as well as explain the muon anomalous magnetic moment. Especially, all these cases except the stau coannihilation can even have low energy fine-tuning measures around 10.

\section{Appendix}
The non-decoupling effect is calculated in terms of Mathematica, which can be found in our previous paper ~\cite{Ding:2015wma}.
The whole process is tedious. So, we just show some key steps:

\begin{enumerate}
\item Getting the scalar potential part of Lagrangian.

  Here, the scalar potential is expressed as a function of $H_u^0$, $H_d^0$, $T_+^0$ and $T_-^0$. 
  In addition to the conventional terms such as $\mu$, $B_{\mu}$ and $m_{\phi}^2$, their quartic term is uniquely determined by the gauge couplings $g_1$ and $g_2$. 
  The general form of scalar potential can be illustrated as follows
    \begin{align}
    V=\mu^2 \phi_i\phi_j+B_{ij}\phi_i\phi_j+A_{ijk}\phi_i\phi_j\phi_k+h.c.+m_i^2\phi_i\phi_i^* +\frac{1}{2}(g_1^2+g_2^2)\phi_i^4
    \end{align}
  where $\phi_i$ stands for the scalar particle that we are interested in. 

\item  Integrating out the massive scalar triplet particles.

  In supersymmetric models, the heavy degrees of freedom can be integrated out through the equations $\partial W/\partial \Phi=0$ due to the F-flatness conditions. After solving these equations, the heavy superfield $\Phi$ can be re-expressed in terms of light superfields. Then substituting the solution into superpotential yields an effective theory with light superfields. In this procedure, supersymmetry is preserved since we integrate out a supermultiplet. However, such an integrating out procedure only affect Higgs mass mildly which means all the heavy superfields are decoupled from the new sector. This strongly motivates us to consider another method where we only integrate out the scalar component of supermultiplet. Thus, we resort to solving the equation $\partial V/\partial \phi=0$ which is called semi-soft supersymmetry breaking. 
  In our model, the solution after taking the limit $m_{T}^2>M_V^2$ are given by
    \begin{align}
    T_+^0&=-\frac{M_V H_u^{2}\lambda}{M_V^2+m_{T_+}^2}\nonumber\\
    T_-^0&=\frac{ H_u(-2H_d\lambda\mu+T_{\lambda}H_u)}{M_V^2+m_{T_-}^2}
    \label{eqn:sol}
    \end{align}
\item Substituting the solution into scalar potential and obtain a new quartic coupling;
    
  After solving the equation, we can substitute the solution in equation~(\ref{eqn:sol}) into original scalar potential. It is easy to find that we can obtain additional quartic coupling
    \begin{align}
    \lambda^2_{eff}=\frac{\lambda^2 m_{T_+}^2}{M_V^2+m_{T_+}^2}
    \end{align}
  So it is clear to us that even the scalar soft mass $m_{T_+}^2$ is set to be very large, the $\lambda_{eff}$ is still non-zero which is called non-decoupling effect. The interesting point is that the large $m_{T_+}^2$ does not appear in the renormalization equation of $m_{H_u}^2$, and thus does not have any effect on naturalness. 
 
\item Solving the tadpole equation, obtaining Higgs mass matrix, and getting Higgs mass eigenvalues.
    
  Even in the effective scalar potential, the tadpole equation $\partial V_{new}/\partial H_u=0$ and $\partial V_{new}/\partial H_d=0$ must be imposed in order to assure the existence of vacuum,    
    \begin{align}
    & -B_{\mu} v_d+m_{H_u}^2 v_u-\frac{1}{8}(g_1^2+g_2^2)v_d^2 v_u+\frac{1}{4}(g_1^2+g_2^2) v_u^3+2\lambda^2 v_u^3-\frac{2M_V^2 v_u^3\lambda^2}{M_V^2+m_{T_+}^2}+v_u\mu^2=0\nonumber\\
    & -B_{\mu} v_u+m_{H_d}^2 v_d-\frac{1}{8}(g_1^2+g_2^2)v_u^2 v_d+\frac{1}{4}(g_1^2+g_2^2) v_d^3+2\lambda^2 v_d^3+v_d\mu^2=0
    \end{align}
  And then we can use the tadpole equations to solve for $m_{H_u}^2$ and $m_{H_d}^2$. After substituting the $m_{H_u}^2$ and $m_{H_d}^2$ into the scalar potential, we can obtain the final form of the scalar potential. Differentiating the scalar potential twice, we can obtain the Higgs mass matrix
    \begin{align}
    M_{hh}=
            \left(
              \begin{array}{cc}
                M_{11} & M_{12} \\
                M_{11} & M_{22} \\
              \end{array}
            \right)
    \end{align}
  where the $M_{ij}$ denotes for the element of matrix
    \begin{align}
    M_{11}&=m_z^2 \cos^2\beta+m_A^2 \sin^2\beta~,~\nonumber\\
    M_{12}&=-m_A^2\cos\beta\sin\beta-m_z^2\cos\beta\sin\beta+2\alpha-\alpha\gamma\cot\beta ~,~\nonumber\\
    M_{21}&=M_{12}~,~\nonumber\\
    M_{22}&=m_A^2\cos^2\beta+m_z^2\sin^2\beta-\left(\frac{2\alpha}{\gamma}\right)+
    2v^2\kappa\lambda^2\sin^2\beta+4\alpha\cot\beta~,~
    \end{align}
  where $\gamma$, $\alpha$ and $\kappa$ are 
    \begin{align}
    \gamma&=\frac{\mu}{T_{\lambda}}~,~ \nonumber\\
    \alpha&=\frac{T_{\lambda}v^2\lambda\mu\sin^2\beta}{M_V^2+m_{T_{-}}^2}~,~\nonumber\\
    \kappa&=\frac{m_{T_{+}}^2}{M_V^2+m_{T_{+}}^2}~.~
    \end{align}
  Notice that $1/\gamma$ and $\alpha$ are vanishing at the limit of small $T_{\lambda}$, there is only one dimensionless parameter $\kappa$ that is relevant to the Higgs mass. After diagonalizing the mass matrix we find there is additional mass contribution to Higgs mass
    \begin{align}
    \delta m_h^2=2v^2\kappa\lambda^2\sin^4\beta= 2 v^2 \lambda_{\text{eff}}^2  \sin^4\beta 
    \end{align}
    
  The main difference between our model and DiracNMSSM comes from the fact that the triplets $T_-$ and $T_+$ can only couple to $H_u H_u$ and $H_d H_d$ respectively in our model rather than singlet $S$ couples to $H_u H_d$ in Ref.~\cite{Lu:2013cta,Kaminska:2014wia}. This is the reason why we get the correction proportional to  $\sin^4\beta$ but not to $\sin^2 2\beta$.
    
\end{enumerate}


\begin{acknowledgments}

This research was supported in part by the Projects 11475238 and 11647601 supported by National Natural Science Foundation of China, and by Key Research Program of Frontier Science, CAS.
The numerical results described in this paper have been obtained via the HPC Cluster of ITP-CAS.

\end{acknowledgments}

\end{document}